\documentclass[twocolumn,prl,aps,showpacs,floatfix,superscriptaddress,preprintnumbers,amsmath,amssymb]{revtex4}
\usepackage{graphicx}
\usepackage{endnotes}
\usepackage{miller,color}

\begin{document}

\newcommand{\todo}[1]{\textbf{\textsc{\textcolor{red}{(TODO: #1)}}}}
\newcommand{\fcs}{Fe$_{1-x}$Co$_{x}$Si}
\newcommand{\mfs}{Mn$_{1-x}$Fe$_{x}$Si}
\newcommand{\mcs}{Mn$_{1-x}$Co$_{x}$Si}
\newcommand{\cso}{Cu$_{2}$OSeO$_{3}$}
\newcommand{\dmdh}{$\mathrm{d}M/\mathrm{d}H$}


\title{Uniaxial pressure dependence of magnetic order in MnSi}

\author{A. Chacon}
\affiliation{Physik-Department, Technische Universit\"at M\"unchen, James-Franck-Str., D-85748 Garching, Germany}

\author{A. Bauer}
\affiliation{Physik-Department, Technische Universit\"at M\"unchen, James-Franck-Str., D-85748 Garching, Germany}

\author{T. Adams}
\affiliation{Physik-Department, Technische Universit\"at M\"unchen, James-Franck-Str., D-85748 Garching, Germany}

\author{F. Rucker}
\affiliation{Physik-Department, Technische Universit\"at M\"unchen, James-Franck-Str., D-85748 Garching, Germany}

\author{G. Brandl}
\affiliation{Physik-Department, Technische Universit\"at M\"unchen, James-Franck-Str., D-85748 Garching, Germany}
\affiliation{Heinz Maier-Leibnitz Zentrum (MLZ), Technische Universit\"at 
M\"unchen, Lichtenbergstr. , D-85748 Garching, Germany}

\author{R. Georgii}
\affiliation{Physik-Department, Technische Universit\"at M\"unchen, James-Franck-Str., D-85748 Garching, Germany}
\affiliation{Heinz Maier-Leibnitz Zentrum (MLZ), Technische Universit\"at 
M\"unchen, Lichtenbergstr. , D-85748 Garching, Germany}

\author{M. Garst}
\affiliation{Institute for Theoretical Physics, Universit\"at zu K\"oln, Z\"ulpicher Str. 77, D-50937 K\"oln, Germany}

\author{C. Pfleiderer}
\affiliation{Physik-Department, Technische Universit\"at M\"unchen, James-Franck-Str., D-85748 Garching, Germany}

\date{\today}

\begin{abstract}
We report comprehensive small angle neutron scattering (SANS) measurements complemented by ac susceptibility data of the helical order, conical phase and skyrmion lattice phase (SLP) in MnSi under uniaxial pressures. For all crystallographic orientations uniaxial pressure favours the phase for which a spatial modulation of the magnetization is closest to the pressure axis. Uniaxial pressures as low as 1\,kbar applied perpendicular to the magnetic field axis enhance the skyrmion lattice phase substantially, whereas the skyrmion lattice phase is suppressed for pressure parallel to the field. Taken together we present quantitative microscopic information how strain couples to magnetic order in the chiral magnet MnSi.
\end{abstract}


\vskip2pc

\maketitle



Skyrmion lattices in chiral magnets attract great interest as an approach to resolve the main limitations of present day spintronics applications.\cite{muhlbauer2009skyrmion,Jonietz,Schultz,Milde,Mochizuki,Romming,Schwarze,pfleiderer2010skyrmion,Yu:2010,seki2012observation,Fert,Tokura}. At the same time skyrmions in chiral magnets receive also great interdisciplinary interest in fields such as soft or nuclear matter \cite{Ackerman:PRE2014,Rho:MPLB2015}, in which their existence and stability has been considered for a long time in terms of generalised elasticity theories. Skyrmions have by now been identified in a wide range of materials classes at temperatures up to 400\,K\cite{muhlbauer2009skyrmion,Yu:2010,seki2012observation,Yu:PNAS2012,Kezsmarki:NatureMaterials,Tokunaga:arXiv}, where systems crystallising with the so-called B20 structure have been studied most extensively (pedagogical introductions and reviews may be found in Ref.\,\cite{muhlbauer2009skyrmion,Jonietz,Schultz,Milde,Schwarze,Tokura}). Yet, one of the most pressing unresolved questions continues to be the microscopic mechanism at the heart of the formation and stability of skyrmions and how to tailor and control their formation and destruction. While the manipulation of skyrmions by spin and magnon currents has been explored   \cite{Jonietz,Schultz,Milde,Mochizuki,Romming}, essentially nothing is known experimentally about the effects of controlled variations of mechanical strain as the most generic theoretical aspect considered for many decades. 

\begin{figure}[t!]
\includegraphics[width=0.45\textwidth]{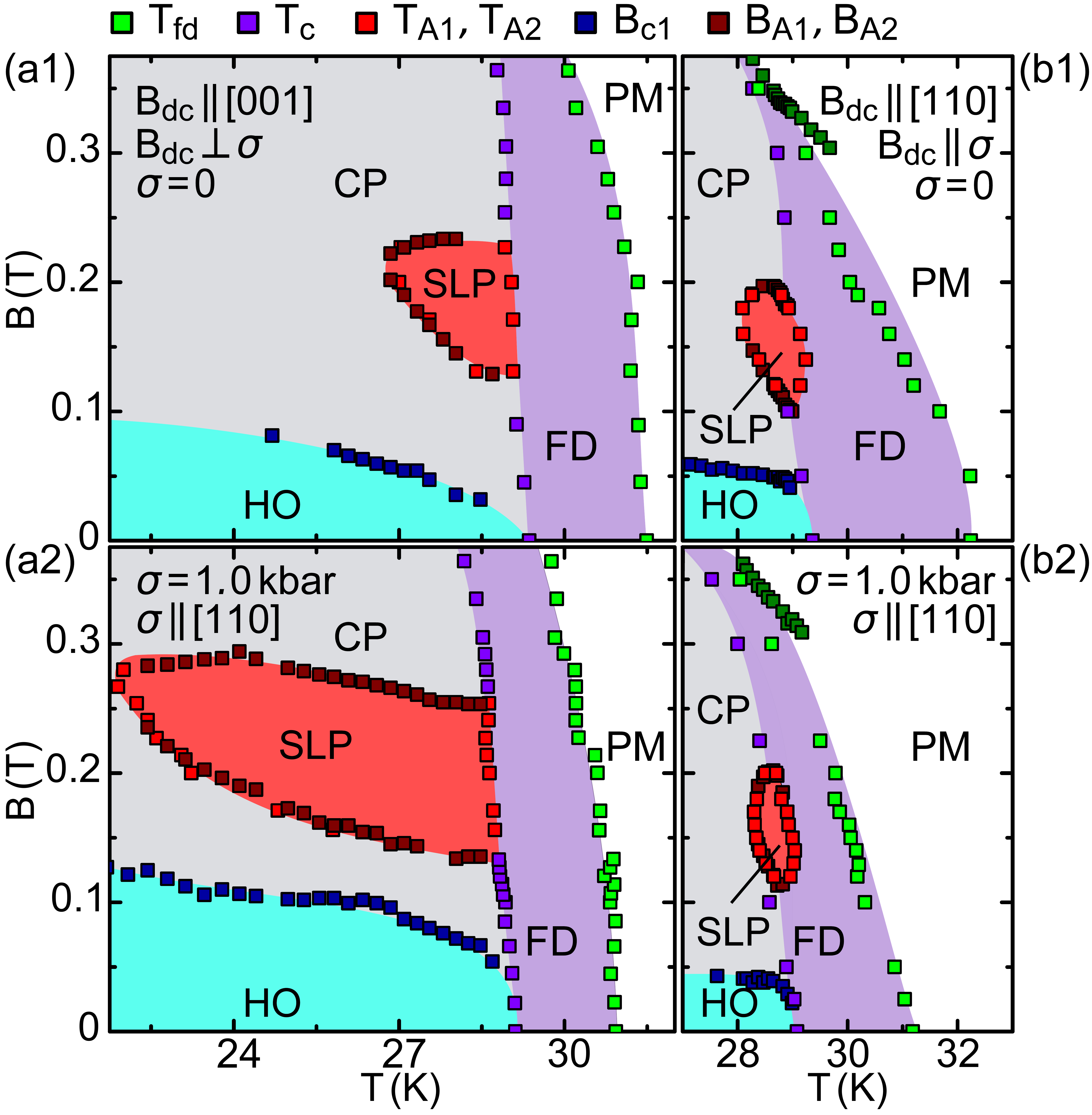}
\caption{(Color online) Typical phase diagrams at ambient and uniaxial pressure for different field directions. HO, CP, SLP, FD and PM correspond to the definitions given in the text. $B_{\rm c2}$ remains unchanged at ca. 600\,mT. (a1) Phase diagram for field along \hkl [001] and ambient pressure. (a2) Phase diagram for field along \hkl [001] \textit{perpendicular} to a uniaxial pressure of 1\,kbar along \hkl [110]. (b1) Phase diagram for field along the \hkl [110] axis and ambient pressure. (b2) Phase diagram for field along \hkl [110] \textit{parallel} to a uniaxial pressure of 1\,kbar along  \hkl [110].
}
\label{figure-4}
\end{figure}

The B20 materials are ideal model systems, as their properties originate from a well-understood hierarchy of energy scales  \cite{Landau:vol8}, comprising of ferromagnetic exchange on the strongest scale, isotropic Dzyaloshinsky-Moriya (DM) spin-orbit-coupling on intermediate scales and higher-order spin-orbit-coupling on the weakest scale. At zero magnetic field helimagnetic order (HO) stabilises below $T_{\rm c}$, where the modulation length varies between $10^2$ and $10^3$\,{\AA}, depending on the specific material. The propagation direction of the modulation is determined by cubic magnetic anisotropies that are fourth order in spin-orbit-coupling.  As illustrated in Fig.\,\ref{figure-4}\,(a1) and (b1) for MnSi, the material selected for our study, fluctuations in the paramagnetic (PM) state assume a strong helimagnetic character just above $T_c$, also referred to as fluctuation-disordered (FD) regime, until the fourth order spin-orbit coupling triggers a fluctuation-induced first order transition \cite{brazovskij1976first,janoschek2013fluctuation,kindervater2014critical}. 

The hierarchy of scales is also at the heart of the magnetic phase diagram shown in Fig.\,\ref{figure-4}\,(a1) and (b1). At $B_{\rm c1}$, determined by the fourth order spin-orbit-coupling terms, a spin-flop transition to the so-called conical phase (CP) occurs \cite{ishikawa1976helical, lebech1989magnetic, adams2012long}. This is followed by the transition to a spin-polarized state (FM) at $B_{\rm c2}$, where $B_{\rm c2}$ reflects the ratio of the FM exchange to DM-interaction. For temperatures just below $T_{\rm c}$ and small fields the skyrmion lattice phase (SLP) is stabilised \cite{muhlbauer2009skyrmion, pfleiderer2010skyrmion,Yu:2010,seki2012observation}. However, as a function of field direction the extent of the skyrmion lattice varies by a factor of two \cite{bauer2012magnetic}.  This has been attributed to the fourth order spin-orbit-coupling terms, whereas the orientation of the skyrmion lattice perpendicular to the field is determined by terms sixth order in spin-orbit-coupling \cite{muhlbauer2009skyrmion}.

The formation of the skyrmion lattice in B20 compounds is dominantly driven by thermal fluctuations \cite{muhlbauer2009skyrmion,BuhrandFritz:PRB}. However, as the temperature range of the skyrmion lattice reflects variations due to higher-order spin-orbit-coupling, a promising route to control the skyrmion lattice appear to be changes of the spin-orbit-coupling. Moreover, the possible existence of skyrmion lattices in chiral magnets was first anticipated theoretically in a  mean-field model for materials with intrinsic or superimposed uniaxial magnetic anisotropy \cite{Bogdanov1989, bogdanov1994thermodynamically}. This has been followed-up by a theoretical study considering uniaxial magnetic anisotropy induced by uniaxial pressure \cite{butenko}. Hence a key question concerns how lattice strain couples to  modulated spin structures in chiral magnets with special interest in the B20 compounds.


Experimentally, the role of large isotropic lattice strain has been studied in hydrostatic pressure experiments in MnSi, FeGe, Fe$_{1-x}$Co$_{x}$Si $(x\!=\!0.1, 0.2)$ and Cu$_2$OSeO$_3$ \cite{pfleiderer1997magnetic, pedrazzini2007metallic, huang2011observation,forthaus2011pressure}. For small pressures $T_{\rm c}$ varies linearily and the magnetic phase diagram remains qualitatively unchanged
\cite{Thessieu:JPCM1997,Ritz:PRB2013,Ritz:Nature2013}. 
 In contrast, the effects of large isotropic lattice strain are still not settled and potentially at the heart of a generic non-Fermi liquid state in three dimensions \cite{pfleiderer2001non, pfleiderer2007non,pedrazzini2007metallic}. This compares with the effects of anisotropic lattice strain, as generated in thin films grown on different substrates \cite{Monchevsky,Marrows,Wiedemann}. However, for the studies reported to date the lattice mismatch with the substrates is prohibitively large and does not allow to establish a straight forward link with the prestine bulk properties. More informative is a recent study of thin bulk samples of FeGe by means of Lorentz TEM \cite{shibata2015large}, which suggests a strong response to small symmetry breaking strains but lacks comprehensive information. Likewise, uniaxial pressure has been used in a study of the helimagnetic state in polycrystalline MnGe  \cite{deutsch2014stress}, which reveals a strong response but does not provide the desired insights. 

In our Letter we report comprehensive small angle neutron scattering (SANS) measurements of MnSi  under uniaxial pressure complemented by some ac susceptibility data \cite{chacon2009,chacon2011}. MnSi belongs to the B20 transition metal compounds, the most extensively studied class of materials exhibiting skyrmion lattices. As our main result we find a remarkably simple behaviour, in which uniaxial pressure imposes a strong easy-axis anisotropy that stabilises the magnetic phase for which a magnetic modulation is closest to the pressure axis, regardless of precise crystallographic orientation. In particular, uniaxial pressures perpendicular to the magnetic field axis enhances the helical order and the skyrmion lattice strongly, whereas both phases are suppressed for pressure parallel to the field as shown in Fig.\,\ref{figure-4}\,(a2) and (b2). 


Our measurements were carried out on four bar-shaped MnSi samples \cite{neubauer2011ultra} for various combinations of pressure and field directions along different crystallographic directions. Uniaxial pressures were generated with a bespoke Helium-activated pressure cell \cite{waffenschmidt1999critical, pfleiderer1997he}. SANS measurements were performed at the MIRA-1 and MIRA-2 beam lines at FRM II \cite{georgii2007mira}. The AC susceptibility $\chi_\textup{ac}$ was measured by means of a bespoke free-standing miniature susceptometer \cite{pfleiderer1997miniature, SOM}. It is helpful to note that uniaxial pressure studies in principle may be subject to inhomogeneous pressure distributions. Further, the skyrmion lattice is sensitive to the precise field distribution inside the sample. The conclusions of our study are purely based on experimental results that are insensitive to these limitations (only applied field values are stated). Further details of the experimental methods and additional data are reported in the supplement \cite{SOM}. 

\begin{figure}[t!]
\includegraphics[width=0.45\textwidth]{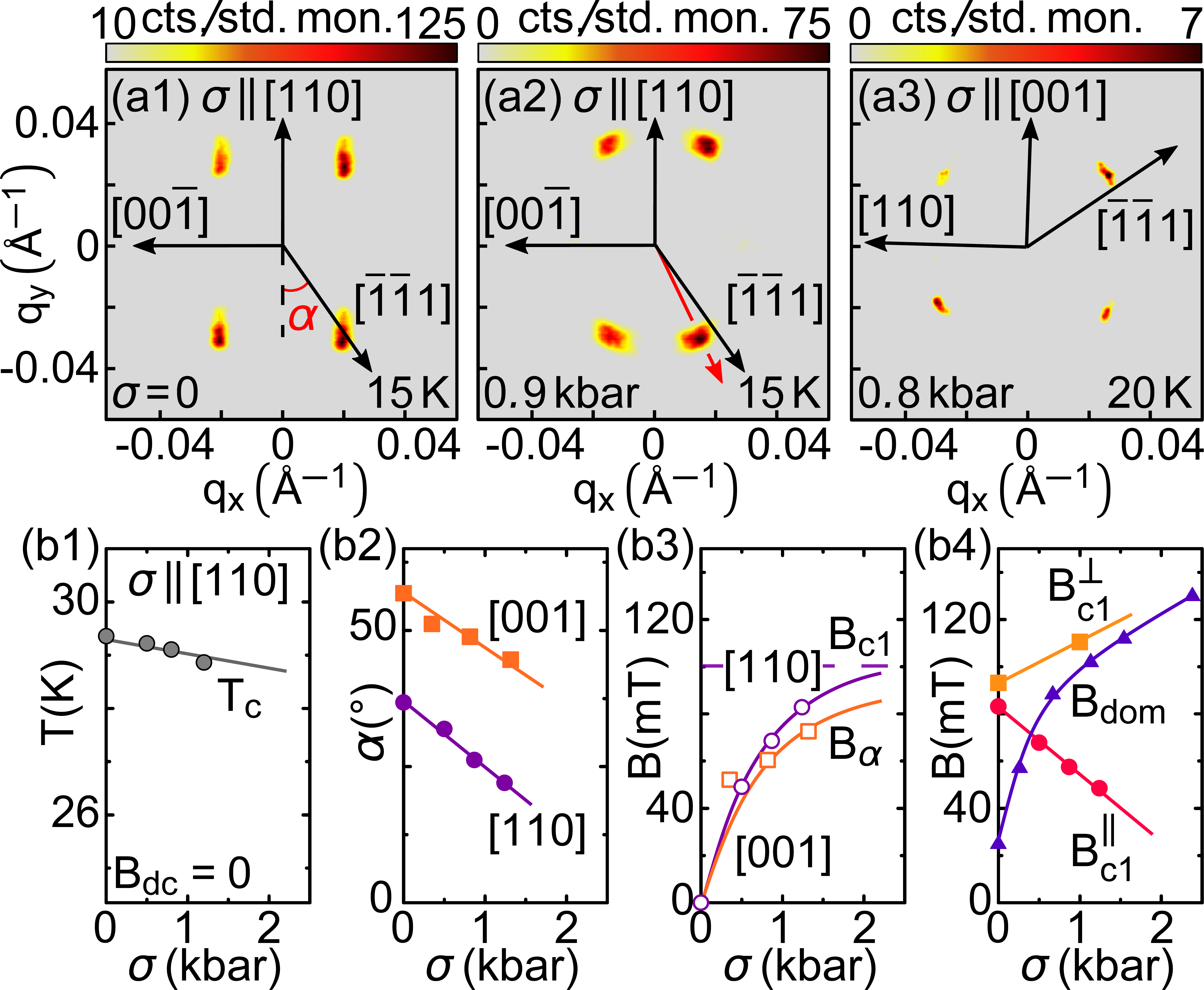}
\caption{(Color online) Typical SANS data of the helical phase of MnSi under uniaxial pressure:
(a1) and (a2): SANS intensity pattern in the helical phase at ambient pressure, and under a uniaxial pressure of 0.9\,kbar parallel to \hkl[110]. 
(a3): SANS intensity pattern in the helical phase and under a uniaxial pressure of 0.8\,kbar parallel to \hkl[001].
(b1), (b2), (b3) and (b4): Uniaxial pressure dependence of $T_{\rm c}$, $B_{\rm c1}$, $B_{\alpha}$ and $B_{\rm dom}$. 
}
\label{figure-1}
\end{figure}

Typical SANS intensity patterns of the helical order are summarised in Fig.\,\ref{figure-1}\,(a1) through (a3). The slight smearing of the intensities may be attributed to pressure inhomogeneities and do not affect the conclusions of our study \cite{SOM}. For $\vec{B}_{\rm dc}=0$ and $\sigma =0$ we observe equal domain populations along the \hkl <111 > axes, as shown in Fig.\,\ref{figure-1}\,(a1). For $\vec{\sigma}$ along \hkl [110] and \hkl [001] the helical propagation direction moves towards the pressure axis, as illustrated in Figs.\,\ref{figure-1}\,(a2) and (a3), respectively. Small systematic tilts away from the scattering plane require a detailed analysis beyond the scope of our study. 

Further, for pressures along \hkl [111] the intensity of the domain populations parallel to the pressure axis increases rapidly until all of the other domains have been completely depopulated \cite{SOM}. Thus, for all crystallographic directions uniaxial pressure imposes a strong easy axis for the modulation direction. However, because we do not observe any evidence for higher-order intensities under uniaxial pressure in the helical state or for any of the other magnetic phases (cf.  Ref.\,\cite{adams:PRL2011} at zero pressure), we conclude that 
uniaxial pressure leaves the magnetic textures basically undistorted and mainly influences their spatial orientation and thermodynamic stability.

It is instructive to distinguish uniaxial pressure-induced \textit{isotropic} strain from the effects of symmetry breaking \textit{anisotropic} strain. Shown in Fig.\,\ref{figure-1}\,(b1) is the zero-field helimagnetic transition temperature, $T_c$, as a function of pressure. The rate of suppression, ${\rm d} T_{\rm c}/{\rm d} \sigma\approx -(0.236\pm0.03)\,{\rm K\, kbar^{-1}}$, corresponds to the measured hydrostatic pressure dependence when assuming an isotropic material, where ${\rm d} T_{\rm c}^{\rm hydro}/{\rm d} p\approx -0.8\,{\rm K\, kbar^{-1}}\approx 3 {\rm d} T_{\rm c}/{\rm d} \sigma$ \cite{pfleiderer1997magnetic}. We conclude that the suppression of $T_{\rm c}$ is dominated by isotropic changes of the unit cell volume.

The effects of symmetry breaking anisotropic strain are quantitatively reflected by the angle $\alpha$ between the helical propagation direction and the pressure axis. Typical data are shown in Fig.\,\ref{figure-1}\,(b2) for pressure along \hkl [110] and \hkl [001], where the linear decrease of $\alpha$ with increasing pressure suggests a linear coupling between strain and helical order. It is possible to express the strength of the pressure-induced anisotropy in terms of the applied magnetic field, $B_{\alpha}$, that causes the same rotation angle $\alpha$ of the helical propagation direction towards the field direction as that generated under pressure. With increasing pressure $B_{\alpha}$ increases rapidly and approaches $B_{\rm c1}$ for \hkl [110] already around $1\,{\rm kbar}$ as shown in Fig.\,\ref{figure-1}\,(b3), where the same behaviour is observed for pressures along \hkl [001] and \hkl [110]. 

\begin{figure}[t!]
\includegraphics[width=0.45\textwidth]{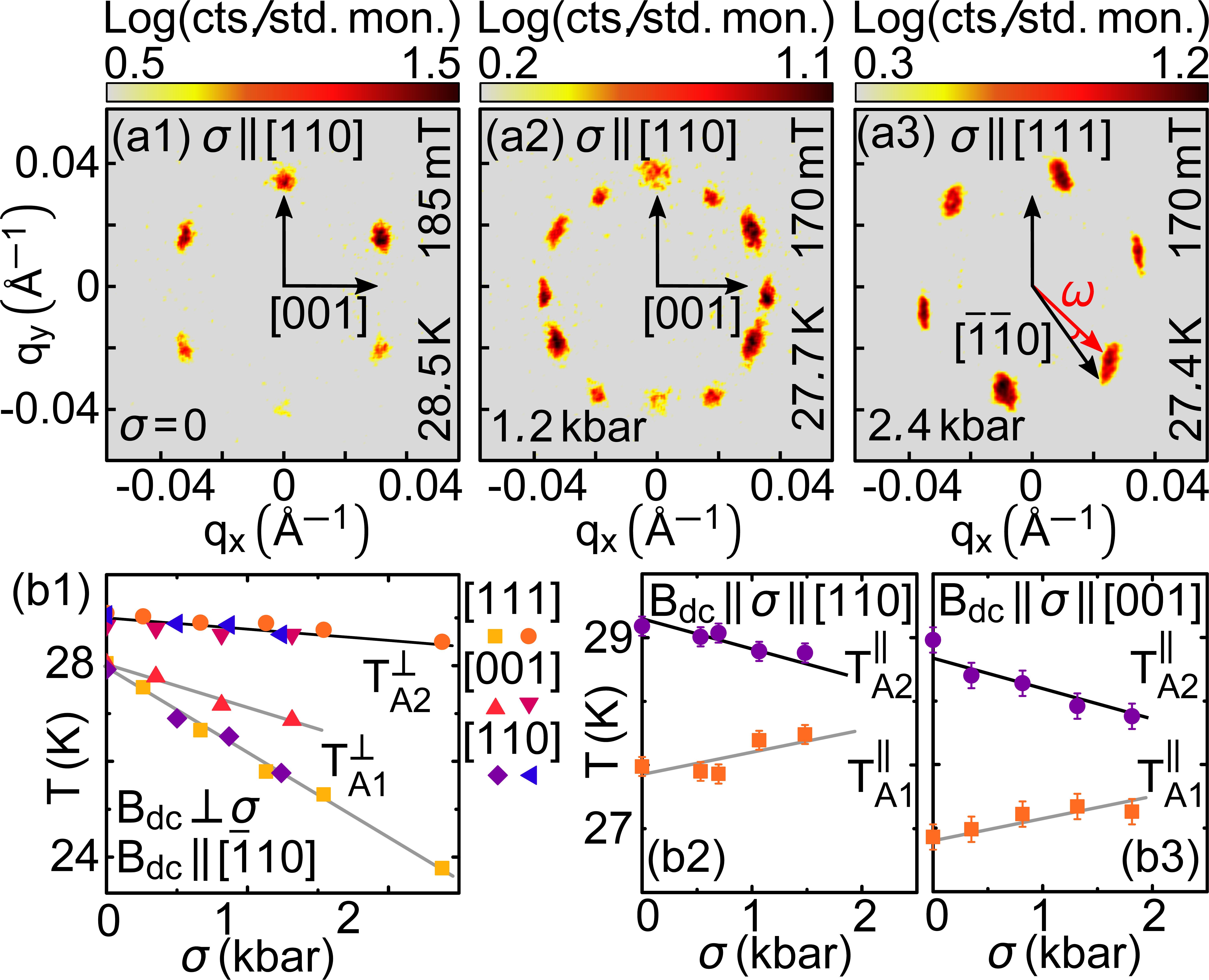}
\caption{(Color online) Typical SANS data of the skyrmion lattice of MnSi under uniaxial pressure.
(a1), (a2): SANS intensity pattern in the SLP with field along \hkl[-110] and $\sigma=0$ as well as $\sigma=1.2\,{\rm kbar}$ parallel \hkl [110]. 
(a3): SANS intensity pattern in the SLP for $\sigma=2.4\,{\rm kbar}$ parallel \hkl [111]. The angle $\omega$ reflects the deviation from the \hkl [110] direction.
(b1), (b2) and (b3): Uniaxial pressure dependence of $T_{\rm A1}$ and $T_{\rm A2}$.
}
\label{figure-2}
\end{figure}



The pressure dependence of  $B_{\rm c1}$ for fields perpendicular and parallel to the pressure axis, denoted $B_{\rm c1}^{\perp}$ and  $B_{\rm c1}^{\parallel}$, respectively are shown in Fig.\,\ref{figure-1}\,(b4) for pressure along \hkl [110] and field parallel \hkl [001] as well as \hkl [110], respectively. There are small differences of $B_{\rm c1}$ at $\sigma=0$ due to the fourth order spin-orbit-coupling terms and $B_{\rm c1}^{\parallel}$ decreases while $B_{\rm c1}^{\perp}$ increases. This is consistent with an increasing uniaxial anisotropy parallel to the pressure axis. To measure the strength of the pressure-induced easy-axis anisotropy for pressure along \hkl [111], we determined the magnetic field, $B_{\rm dom}$, applied perpendicular to the pressure axis. $B_{\rm dom}$ is defined as the field at which the domains along the pressure axis become completely depopulated. As a function of pressure $B_{\rm dom}$ increases rapidly and approaches $B_{\rm c1}^{\perp}$ from below for pressures of the order of $1\,{\rm kbar}$ as shown in Fig.\,\ref{figure-1}\,(b4). Taken together the pressure-induced anisotropy exceeds the cubic anisotropy and dominates above a few kbars, regardless along which crystallographic axes pressure is applied. 



In contrast to the response of the helical order, key aspects of the scattering pattern of the skyrmion lattice remain qualitatively unchanged for all pressure and field combinations, in marked contrast with recent TEM work on thin FeGe samples \cite{deutsch2014stress}. This is illustrated in Figs.\,\ref{figure-2}\,(a1) through (a3) for field parallel \hkl [-110]. We begin in Fig.\,\ref{figure-2}\,(a1) with typical data for $\sigma =0$, where the usual sixfold pattern is observed perpendicular to the field. One pair of Bragg spots is aligned along \hkl [110]. Small differences of the spot intensities arise from an incomplete rocking scan. The changes of the pattern in the SLP for pressures perpendicular to the applied field may be summarised as follows (see \cite{SOM} for further data). First, for pressure along \hkl [110] a second domain population appears in field sweeps after zero-field cooling, as shown in Fig.\,\ref{figure-2}\,(a2). The field and temperature regimes of the two domain populations differ slightly \cite{SOM}. Second, for pressures along \hkl [111] the pattern rotates towards the pressure axis \cite{SOM}, cf. Fig.\,\ref{figure-2}\,(a3). Third, for pressures along \hkl [001], i.e., the hard axis within the skyrmion lattice plane, the skyrmion lattice remains aligned along \hkl [110] for all pressures without evidence for additional domains \cite{SOM}. 

Our observations on the skyrmion lattice may be explained within an effective description for the orientation angle $\omega$ quantifying the deviation of one of the Bragg spots from the \hkl <110> direction in terms of the potential  
$\mathcal{V}(\omega) = - V(\sigma) \cos(6 (\omega- \omega_0(\sigma)))$
with pressure dependent coefficients $V(\sigma)$ and $\omega_0(\sigma)$ (cf Ref.\,\cite{muhlbauer2009skyrmion}). For zero pressure, $\sigma=0$, and a magnetic field along $[\bar{1}10]$, the 180$^\circ$ rotation symmetry around the cubic \hkl [001] axes of the tetrahedral point group ensures that the angle $\omega_0(0) = 0$; furthermore $V(0) > 0$ so that the potential is minimized for $\omega = 0$ consistent with a Bragg spot along \hkl [110]. 

\begin{figure}[t!]
\includegraphics[width=0.45\textwidth]{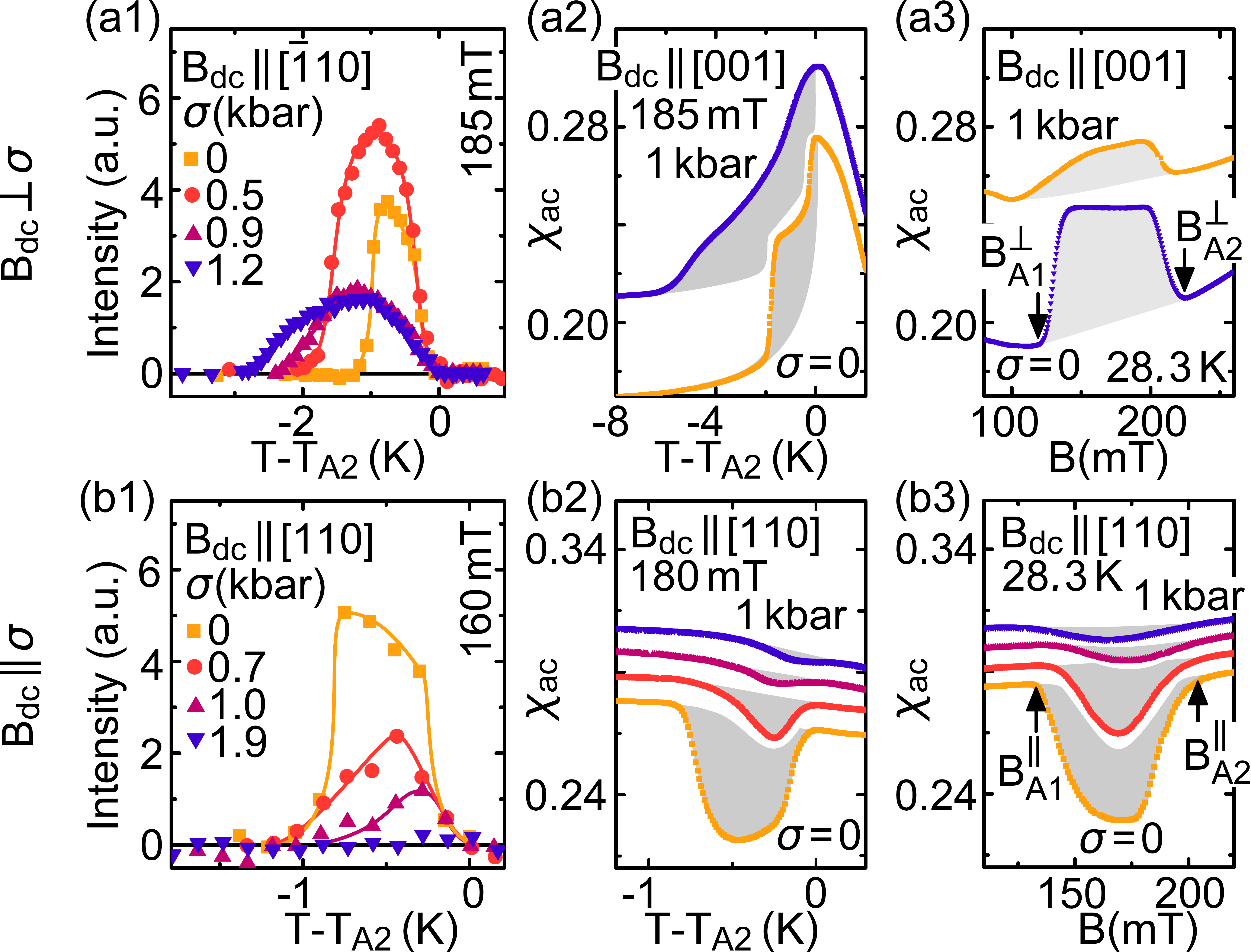}
\caption{(Color online) Typical temperature and field dependence of SANS intensity and ac susceptibility of MnSi under various uniaxial pressures. Also shown are characteristic fields and temperatures inferred from these data. 
(a1) and (b1): Typical temperature and field dependencies of SANS peak intensities.
(a2), (a3), (b2) and (b3): Typical ac susceptibility data as a function of temperature or magnetic field. 
}
\label{figure-3}
\end{figure}



Uniaxial pressure $\vec \sigma$ along \hkl [110] or \hkl [001] leaves the 180$^\circ$ rotation symmetry around \hkl [001] intact as $\vec \sigma$ is a director, i.e., for these directions $\omega_0(\sigma) = 0$ and only the coefficient $V(\sigma)$ varies. Our observations imply that for the pressures along \hkl [001] the coefficient $V(\sigma)$ remains positive and the pattern is unchanged. In contrast, for pressures along \hkl [110], $V(\sigma)$ decreases and changes sign at a critical pressure $\sigma_{\rm c}\sim 1$ kbar, favouring a domain with $\omega = 30^\circ$ above $\sigma_{\rm c}$. Inhomogeneities in the pressure distribution within the sample lead to the coexistence of domains with different orientations causing the 12-fold symmetry of the SANS pattern in Fig.\ref{figure-2}\,(a2). Finally, for pressures along \hkl [111] the residual 180$^\circ$ symmetry is broken and $\omega_0(\sigma)$ becomes finite resulting in a smooth increase of $\omega$ with increasing $\sigma$ and a drift of the Bragg spot away from \hkl [110] (Fig.\ref{figure-2}\,(a3)).

As an additional effect not observed so far we anticipate that uniaxial pressure also couples to the normal vector of the skyrmion lattice plane inducing a small tilt of a few degrees away from the magnetic field axis. A similar tilt of the skyrmion plane due to cubic anisotropies has been recently resolved as a function of the field orientation in a very careful study on a spherical sample at $\sigma = 0$ \cite{adams2015}.



The phase boundary of the skyrmion lattice was inferred from temperature and field sweeps, where typical SANS data are shown in Fig.\,\ref{figure-3}\,(a1) and (b1). For field perpendicular to the pressure axis the temperature range of the skyrmion lattice increases rapidly, whereas the temperature range is essentially unchanged for field parallel to the pressure axis and the intensity disappears around 2\,kbar. The pressure dependence of the transiton temperatures in ($T_{\rm A1}$) and out ($T_{\rm A2}$) of the skyrmion lattice phase ($T_{\rm A1}$ and $T_{\rm A2}$ respectively) are summarized in Figs. 3(b1) through (b3). For pressures along \hkl [111], \hkl [110] or \hkl [001], and field along \hkl [-110] perpendicular to the pressure axis, the transition temperature to the paramagnetic state, $T_{\rm A2}$, decreases with ${\rm d} T_{\rm A2}/{\rm d} \sigma\approx -(0.236\pm0.03)\,{\rm K\, kbar^{-1}}$ as shown in Fig.\,\ref{figure-2}\,(b1). We note that in Figs.\,\ref{figure-2}\,(b2) and (b3) the slope of the line guiding $T_{\rm A2}$ are the same as determined from a fit of $T_{\rm A2}$ in panel (b1). Taken together these data imply that $T_c$ and $T_{A2}$ are dominated by the trace of the induced strain tensor, which in a cubic crystal is independent of the orientation of the applied uniaxial stress.

The transition at $T_{\rm A1}$ decreases rapidly with increasing pressure, ${\rm d} T_{\rm A1}/{\rm d} \sigma\approx -(2\pm0.05)\,{\rm K\, kbar^{-1}}$ and ${\rm d} T_{\rm A1}/{\rm d} \sigma\approx -(1\pm0.05)\,{\rm K\, kbar^{-1}}$ for pressures along \hkl [111] and \hkl [110], and \hkl [001], respectively. Thus, the rapid decrease of $T_{\rm A1}$, notably the balance between skyrmion lattice and conical phase, dominates how uniaxial pressure perpendicular to the applied field enhances the skyrmion lattice. This compares in particular with the pressure dependence of $T_{\rm A1}$ and $T_{\rm A2}$ for field parallel to the pressure axis, where very weak changes are observed as shown in Figs.\,\ref{figure-2}\,(b2) and (b3). 

The ac susceptibility, $\chi_{ac}$, shown in Figs.\,\ref{figure-3}\,(a2), (a3), (b2) and (b3), was measured to obtain further information on the phase boundaries. The AC excitation  was always oriented parallel to the pressure axis. Thus, Figs.\,\ref{figure-3}\,(a2) and (a3) show $\chi_{ac}$ \textit{transverse} to the applied field. In contrast to the well-understood reduction of $\chi_{ac}$ when the excitation is longitudinal to the applied field \cite{bauer2012magnetic}, $\chi_{ac}$ is enhanced under transverse excitation.

In summary, our data show in Fig.\,\ref{figure-4} typical magnetic phase diagrams, for field parallel to \hkl [001] and \hkl [110], respectively. For pressures applied perpendicular to the field, the extend of the skyrmion lattice phase is strongly enhanced due to changes of $T_{\rm A1}$ (Fig.\,\ref{figure-4}\,(a2)). In contrast, under pressures parallel to the magnetic field direction, the temperature range of the skyrmion lattice decreases only slowly until the skyrmion lattice vanishes (Fig.\,\ref{figure-4}\,(b1) and (b2)). This is consistent with the mean-field prediction of Ref.~\cite{butenko} that a uniaxial tensile stress along the magnetic field stabilizes the SLP. The combination of different field and pressure directions suggests strongly, that our observation does not depend on the precise crystallographic orientation of the sample with respect to the pressure and field direction.


Our study reveals a remarkably simple relationship between uniaxial pressure-induced lattice strain and the magnetic order in the B20 compound MnSi, even though uniaxial pressure breaks the symmetries of the crystal structure. To leading order an easy-axis anisotropy for the modulation direction along the pressure axis is generated, regardless of the crystallographic orientation. This anisotropy becomes remarkably strong already around $\sim1\,{\rm kbar}$. The boundary to the paramagnetic state is dominated by  isotropic strain components. In contrast, amongst the ordered phases, pressure stabilises the magnetic order, for which a modulation direction is closest to the pressure axis. This explains the pressure dependence of $B_{\rm c1}$ as well as the pressure dependence of the phase boundaries separating the skyrmion lattice and conical phase, where the latter increases strongly for pressures perpendicular to the field direction and thus within the skyrmion lattice plane. 

The measurements presented here bridge the gap between the present-day understanding of small isotropic and large anisotropic lattice strains in MnSi as a prototypical model system supporting itinerant helimagnetism and a skyrmion lattice phase, where the latter has been explored in attempts to stabilise a skyrmion lattice in thin films. In this context our results provide the basis for a complete quantitative understanding how to create or destroy skyrmions in nano-scale systems.


We wish to thank P. B\"oni, M. Halder, H. Kolb, S.\ Mayr, J. Neuhaus, J. Peters, M. Pfaller, A. Rosch, K. Seeman, R. Schwikowski, and the team of FRM II, for fruitful discussions and assistance with the experiments. Financial support through DFG TRR80, DFG FOR960, and ERC advanced grant (291079, TOPFIT) is gratefully acknowledged. A.C., A.B., T.A.,  F.R., and G.B. acknowledge financial support through the TUM graduate school.

Note added: We have recently become aware of a related publication \cite{nii2015}, that covers, however, a  much smaller parameter range than our study. 




\end{document}


\newcommand{\todo}[1]{\textbf{\textsc{\textcolor{red}{(TODO: #1)}}}}
\newcommand{\OLD}[1]{{\tiny {\bf old:} #1 }}
\newcommand{\NEW}[1]{{ \it #1 }}
\renewcommand{\vec}[1]{\boldsymbol{#1}}
\newcommand{\w}{\omega}

\newcommand{\fcs}{Fe$_{1-x}$Co$_{x}$Si}
\newcommand{\mfs}{Mn$_{1-x}$Fe$_{x}$Si}
\newcommand{\mcs}{Mn$_{1-x}$Co$_{x}$Si}
\newcommand{\cso}{Cu$_{2}$OSeO$_{3}$}

\newcommand{\rxx}{$\rho_{\rm xx}$}
\newcommand{\rxy}{$\rho_{\rm xy}$}
\newcommand{\rxyt}{$\rho_{\rm xy}^{\rm top}$}
\newcommand{\Drxyt}{$\Delta\rho_{\rm xy}^{\rm top}$}
\newcommand{\Sxy}{$\sigma_{\rm xy}$}
\newcommand{\Sxya}{$\sigma_{\rm xy}^A$}

\newcommand{\bco}{$B_{\rm c1}$}
\newcommand{\bct}{$B_{\rm c2}$}
\newcommand{\bao}{$B_{\rm A1}$}
\newcommand{\bat}{$B_{\rm A2}$}
\newcommand{\beff}{$B^{\rm eff}$}

\newcommand{\btr}{$B^{\rm tr}$}

\newcommand{\tc}{$T_{\rm c}$}
\newcommand{\ttr}{$T_{\rm tr}$}

\newcommand{\mb}{$\mu_0\,M/B$}
\newcommand{\dmdb}{$\mu_0\,\mathrm{d}M/\mathrm{d}B$}
\newcommand{\ddmddb}{$\mathrm{\mu_0\Delta}M/\mathrm{\Delta}B$}
\newcommand{\cm}{$\chi_{\rm M}$}
\newcommand{\cac}{$\chi_{\rm ac}$}
\newcommand{\rechi}{${\rm Re}\,\chi_{\rm ac}$}
\newcommand{\imchi}{${\rm Im}\,\chi_{\rm ac}$}

\newcommand{\ozz}{$\langle100\rangle$}
\newcommand{\ooz}{$\langle110\rangle$}
\newcommand{\ooo}{$\langle111\rangle$}
\newcommand{\too}{$\langle211\rangle$}

\makeatletter
\renewcommand{\thefigure}{S\@arabic\c@figure}
\makeatother

\renewcommand{\floatpagefraction}{0.5}
\setcounter{secnumdepth}{2}

\title{Supplementary Material for:\\ 
Uniaxial pressure dependence of magnetic order in MnSi}

\author{A. Chacon}
\affiliation{Physik-Department, Technische Universit\"at M\"unchen, D-85748 Garching, Germany}

\author{A. Bauer}
\affiliation{Physik-Department, Technische Universit\"at M\"unchen, D-85748 Garching, Germany}

\author{T. Adams}
\affiliation{Physik-Department, Technische Universit\"at M\"unchen, D-85748 Garching, Germany}

\author{F. Rucker}
\affiliation{Physik-Department, Technische Universit\"at M\"unchen, D-85748 Garching, Germany}

\author{G. Brandl}
\affiliation{Physik-Department, Technische Universit\"at M\"unchen, D-85748 Garching, Germany}
\affiliation{Heinz Maier-Leibnitz Zentrum (MLZ), Technische Universit\"at 
M\"unchen, Lichtenbergstr. , D-85748 Garching, Germany}

\author{R. Georgii}
\affiliation{Physik-Department, Technische Universit\"at M\"unchen, D-85748 Garching, Germany}
\affiliation{Heinz Maier-Leibnitz Zentrum (MLZ), Technische Universit\"at 
M\"unchen, D-85748 Garching, Germany}

\author{M. Garst}
\affiliation{Institute for Theoretical Physics, Universit\"at zu K\"oln, Z\"ulpicher Str. 77, D-50937 K\"oln, Germany}

\author{C. Pfleiderer}
\affiliation{Physik-Department, Technische Universit\"at M\"unchen, D-85748 Garching, Germany}

\date{\today}

\begin{abstract}
In this supplement we present detailed information on the experimental methods, a table summarising all measurements performed. We also present additional experimental data that complement the main results reported in the main text of the paper. 
\end{abstract}


\vskip2pc

\maketitle


\section{Table of Abbreviations}

\begin{table*}[hb]
\caption{Abbreviations used in the main text and supplement and their meaning.}
\begin{tabular}{|l|l|}
\hline
Abbreviation & Meaning \\
\hline
SLP & Skyrmion Lattice Phase \\
SOC & Spin Orbit Coupling \\
HO & Helical Order \\
FM & Spin Polarized State \\
CP & Conical Phase  \\
FD & Fluctuation-Disordered Regime \\
DM & Dzyaloshinsky-Moriya \\
SANS & Small Angle neutron Scattering \\
$T_{\rm c}$ & Transition Temperature ($B\!=\!0$) \\
$B_{\rm c1}$ & Transition field from helical to conical phase \\
$B_{\rm c2}$  & Transition field from conical phase to spin polarized state \\
$B_\alpha$ & Equivalent field for reorientation by $\alpha$ \\
$B_{\rm dom}$ & Transition field for depopulation between helical domains \\
$T_{\rm c}^{\rm hydro}$ & Transition temperature under hydrostatic pressure \\
$B_{\rm c1}^{\parallel}$ & $B_{\rm c1}$ for uniaxial pressures parallel to the applied magnetic field \\
$B_{\rm c1}^{\perp}$ & $B_{\rm c1}$ for uniaxial pressure perpendicular to the applied magnetic field \\
$B_{\rm dc}$ & External magnetic field \\
$B_{\rm A1}$ & Transition field from conical phase to the skyrmion lattice phase \\
$B_{\rm A2}$ & Transition field from the skyrmion lattice phase to the conical phase \\
$T_{\rm A1}$ & Transition temperature from conical phase to the skyrmion lattice phase \\
$T_{\rm A2}$ & Transition temperature from SLP to the fluctuation disordered regime \\
\hline
\end{tabular}
\label{abbtab}
\end{table*}

\section{Experimental Methods}

\subsection{Sample Preparation}
 
For our study four MnSi samples were prepared from optically float-zoned high-quality single-crystal ingots~\cite{neubauer2011ultra}. Three samples had a size of $5\times 1\times 1\,\textup{mm}^3$ and one sample had a size of $10\times 2\times 2\,\textup{mm}^3$). The samples were oriented by means of x-ray Laue diffraction such that a long axis was parallel to $\langle 100 \rangle$, $\langle 110 \rangle$ and $\langle 111 \rangle$. The larger sample was oriented to have the largest axis along $\langle 110 \rangle$. Uniaxial pressure was applied along the long axis of the samples. The sample quality is comparable to those measured in previous studies \cite{bauer2012magnetic}. Shown in table\,\ref{table} is a summary of all samples studied including the pressure and field directions as well as the experimental probes used.

\subsection{Uniaxial Pressure Cell} 

Uniaxial pressures were generated with a bespoke pressure cell based on a He-activated bellows system \cite{waffenschmidt1999critical, pfleiderer1997he}. The body of the pressure cell was made of Cu:Be and Aluminum AW7077 to combine structural stability with low neutron absorption \cite{chacon2009,chacon2011}. The applied force was measured with a piezo sensor (Kistler 9001A ) and the effective cross-section calibrated from high precision measurements of the gas pressure as measured with a Siemens Diptron 3 Digital Manometer. During the experiments the system remained constantly connected to a 10\,liter gas dump, reducing pressure variations during temperature sweeps to well below 3\,\%.

\subsection{Small Angle Neutron Scattering}
Small angle neutron scattering (SANS) measurements were performed at the MIRA-1 and MIRA-2 beam lines at FRM II \cite{georgii2007mira}. At MIRA-1 the neutron beam was collimated with an aperture, $5\times 5\,\textup{mm}^2$, situated $\sim1.5\,\textup{m}$ before the sample and a second aperture, $5\times5\,\textup{mm}^2$, placed $\sim\! 0.5\,\textup{m}$ in front of the sample. Scattered neutrons were recorded using a delayline $^3$He area detector, 20x20\,$\textup{cm}^2$, with a spatial resolution of 2x2\,$\textup{mm}^2$ placed 0.85\,m behind the sample. At MIRA-2 the first aperture had a size of $2\times 3\,\textup{mm}^2$ and the second had a size of $1\times 1.4\,\textup{mm}^2$. They were placed at the same locations as for MIRA-1, with a CASCADE detector \cite{klein2011cascade} placed $\sim\!2.3\,\textup{m}$ behind the sample. A magnetic field up $250\,\textup{mT}$ was generated with a bespoke set of water-cooled Cu solenoids in a Helmholtz configuration \cite{mühlbauer2005}. Magnetic fields could be applied either parallel or perpendicular to the neutron beam. 

\begin{table*}
\caption{Summary of all sample dimensions, sample orientations, field orientations, SANS configurations (instrument, wavelength and type of detector) as well as measurements performed in the context of this study. SANS: small angle neutron scattering, ACS: ac susceptibility}
\begin{tabular}{|c|c|c|c|c|c|c|c|c|c|}
\hline
sample & size & $\vec{\sigma}$ & $\vec{B}_\textup{dc}$ & $\vec{B}_\textup{ac}$ & $\vec{B}_\textup{dc}\,\textup{vs}\,\vec{\sigma}$ & measurement & beam line & $\lambda\,(\textup{\AA})$ & detector \\
 & &  & & (always $\parallel\sigma$) & & & & &  \\
\hline
S1 & 10x2x2 & \hkl [110] & \hkl [-110] & NA & $\perp$ & SANS & MIRA-1 & 10.4 & He$^3$ PSD\\
S1 & 10x2x2 & \hkl [110] & \hkl [001] & \hkl [110] & $\perp$ & ACS & & &\\
S1 & 10x2x2 & \hkl [110] & \hkl [110] & \hkl [110] & $\parallel$ & ACS & & &\\
S2 & 5x1x1 & \hkl [110] & \hkl [-110] & NA & $\parallel$ & SANS & MIRA-2 & 5 & CASCADE \\
S3 & 5x1x1 & \hkl [001] & \hkl [-110] & NA & $\perp$ & SANS & MIRA-2 & 4.5 & CASCADE\\
S3 & 5x1x1 & \hkl [001] & \hkl [001] & NA & $\parallel$ & SANS & MIRA-2 & 4.5 & CASCADE\\
S4 & 5x1x1 & \hkl [111] & \hkl [-110] & NA & $\perp$ & SANS & MIRA-2 & 4.5 & CASCADE\\
\hline
\end{tabular}
\label{table}
\end{table*}

\subsection{AC Susceptibiity}
The AC susceptibility, $\chi_\textup{ac}$, was measured in the same basic set up used for the SANS studies by means of a bespoke free-standing miniature susceptometer \cite{pfleiderer1997miniature}. The secondaries of the susceptometer were wound concentrically on top of each other while the primary surrounded the pressure cell on the outside. The inner diameter of the secondaries was $3.1\,\textup{mm}$ with a height of $5.5\,\textup{mm}$. AC susceptibility data were recorded at typical excitation fields of $\sim1\,\textup{mT}$ and excitations frequencies of $900\,\textup{Hz}$ oriented parallel to the axis of the uniaxial pressure cell. For measurements up to a few Tesla a cryogen free superconducting magnet was used. The susceptibility data was calibrated against data recorded with a Quantum Design physical properties measurement system. 

\subsection{Field and Pressure Configurations}
The SANS and $\chi_\textup{ac}$ measurements were carried out for two configurations of the uniaxial pressure axis and magnetic field direction. Notably, in the first configuration the applied field was perpendicular to the uniaxial pressure axis, $\vec{B}_{\rm dc}\perp \vec{\sigma}$, and parallel to the incident neutron beam. This geometry allowed to examine the helical order and the full diffraction pattern of the skyrmion lattice. In the second configuration the applied field was oriented parallel to the uniaxial pressure axis, $\vec{B}_{\rm dc}\parallel \vec{\sigma}$ and perpendicular to the incident neutron beam. This geometry allowed to track the helical and the conical order, as well as scattering intensity of the skyrmion lattice perpendicular to the field. Excellent alignment of the samples within the pressure cell was confirmed by the location of the helical intensity pattern in zero magnetic field.

\begin{figure}
\includegraphics[width=0.9\linewidth,clip=]{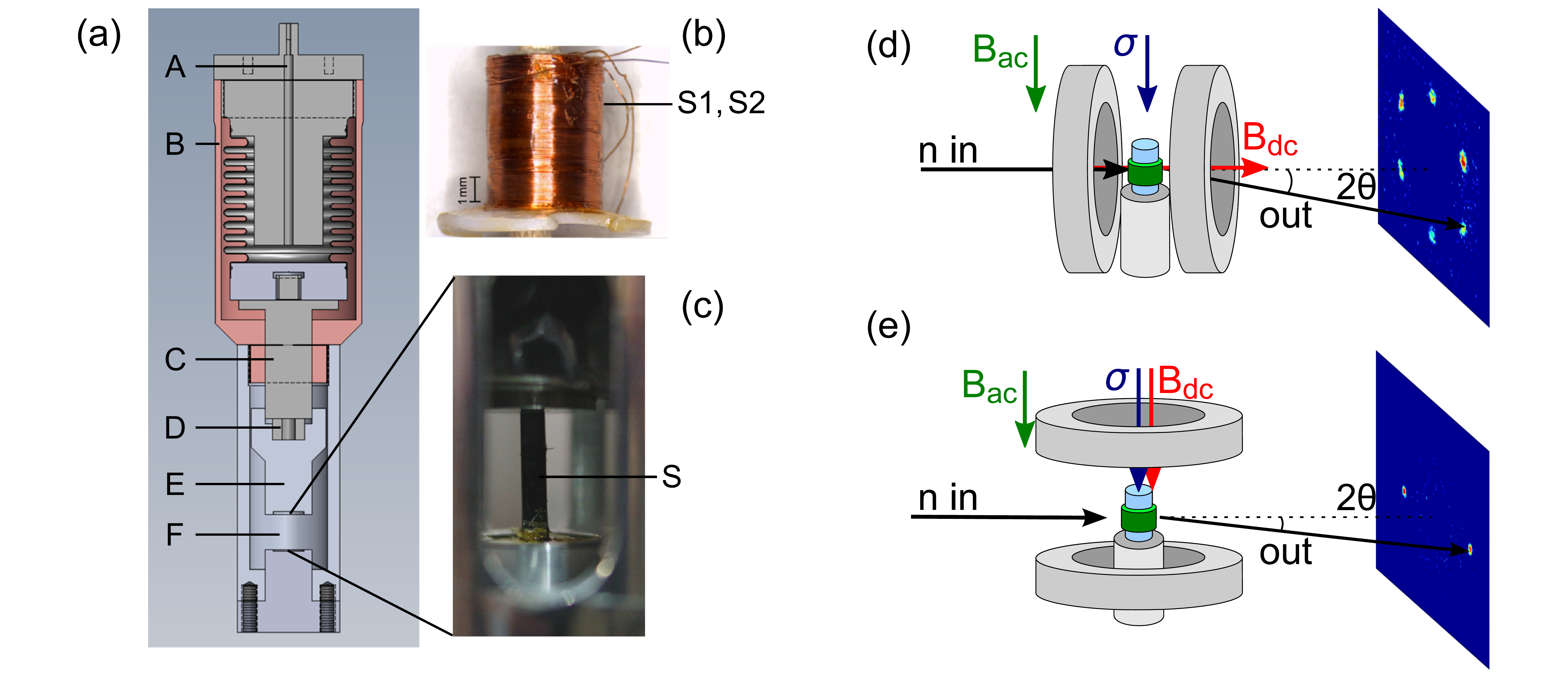}
\caption{(Colour online) Schematic depiction of the uniaxial pressure cell and the SANS set up. (a) Cut-away view of the uniaxial pressure cell, which is based on He-activated bellows. The pressure comprises of \textbf{A} Helium gas capillary, \textbf{B} Cu:Be cage, \textbf{C} Force transmission, \textbf{D} Piezo Sensor, \textbf{E} Uniaxial pressure piston, and \textbf{F} Sample space. (b) Photograph of the pair of secondary detection coils of the ac susceptometer. (c) Photograph of the sample as mounted in the pressure cell for SANS measurements. (d) SANS configuration 1: The applied magnetic field $\vec{B}_{\rm dc}$ is perpendicular to both $\sigma$ and the ac excitation field $\vec{B}_{\rm ac}$. (e) SANS configuration 2:  The applied magnetic field $\vec{B}_{\rm dc}$ is parallel to the pressure axis and the ac excitation field $\vec{B}_{\rm ac}$.
}
\label{Fig_S1}
\end{figure}

\subsection{Pressure Homogeneity}

Uniaxial pressure experiments are sensitive to three main sources of experimental uncertainties: (i) friction, (ii) misalignment, (iii) mechanical damage of the sample. In the following we discuss these aspects and their relevance to our experiments in further detail.

Friction between force transmitting pistons and sample cause a barreling of the sample. Finite element analysis establishes that finite friction leads to variations of the uniaxial pressures across the sample. They are largest in the center of the sample and smallest on the surface. The effects of these redistributions have been noticed in the resistivity and susceptibility of a superconducting transition as described in Ref.\,\cite{pfleiderer1997he}. The effects of friction are greatly reduced for samples with a large aspect ratio, where the height is much larger than the width or depth \cite{Devoile:DEA1996}. The effects of friction may be reduced by means of a carefully selected material placed at the interface between the pistons and the sample surface.

To minimise the effects of friction in our experiments we have studied samples with an aspect ratio of at least five. The sample were separated with cigarette paper from the force transmitting pistons (this approach is based on experience). Our SANS data displays some broadening of the magnetic Bragg spots in the presence of uniaxial pressures. These suggest the presence of some inhomogeneities, albeit they appear to be weak.

Incorrect misalignment may generate shear stresses in the sample beyond the effects of uniaxial pressure. In our experiments we have confirmed parallel alignment better than $1^{\circ}$ between the pressure axis and the crystallographic axis in two steps. First, without pressure the diffraction pattern seen in our SANS measurements provide very accurate information on the alignment (note that the neutron scattering instrument and the sample environment are already aligned very accurately). Second, during our SANS experiments the diffraction pattern in the different magnetic phases clearly showed that the alignment of the sample remained unchanged. 

Last but not least we rule out mechanical damage to the samples during the experiments base on an excellent reproducibility of our observations in repeated loading cycles in our measurements on samples S1, S2, S3. This confirms that our results reflect the effects of uniaxial pressure.


\section{Further Experimental Data}

In this section we present further data recorded in our study. These data are mentioned in the main text and represent secondary information that back up the main observations reported in our paper.

\subsection{Helical Phase}

\begin{figure}[h!]
\includegraphics[width=0.8\linewidth,clip=]{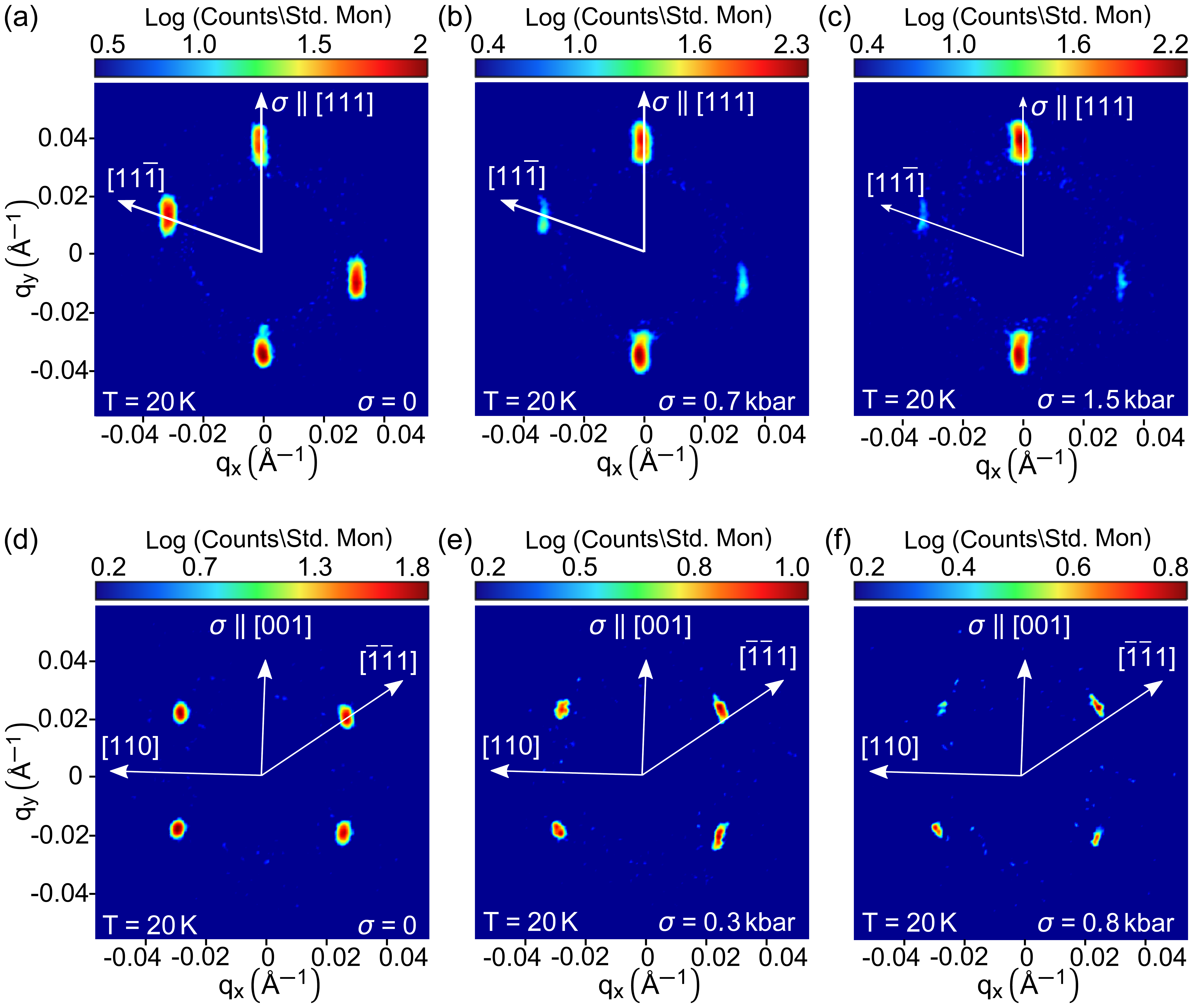}
\caption{(Colour online) Intensity distribution in the helical state for $\vec{B}_{\rm dc}=0$ and various uniaxial pressures along a \hkl [111] or \hkl [001] (the pressure axis was here the vertical direction). Panels (a) through (c): Intensity distribution for various pressures along \hkl [111]. Panels (d) through (f): Intensity distribution for various pressures along \hkl [001].}
\label{Fig_S2}
\end{figure}

Shown in Fig.\,\ref{Fig_S2} are typical intensity patterns of the HO for uniaxial pressures along \hkl [111] or \hkl [001]. For pressure along \hkl [111] we observe equal domain populations at ambient pressure, as shown in Fig.\,\ref{Fig_S2}\,(a), and a change of domain population at high pressures where the domains away from the pressure axis become depopulated (Fig.\,\ref{Fig_S2}\,(b) and (c)).  The direction of the HO is, in contrast, not affected for pressure along \hkl [111]. 

For pressure along \hkl [001] the situation evolves rather differently, as shown in Figs.\,\ref{Fig_S2}\,(d) through (f). Here the most prominent feature of the intensity patterns is a rotation of the direction of the HO towards the pressure axis. However, closer inspection reveals also changes of intensity. While a comprehensive search for the precise fate of the intensity was beyond the scope of our study, a few rocking scans suggest that the propagation direction of the HO, as a function of pressure, does not approach the pressure axis on the shortest possible trajectory. Instead, even though the intensity clearly moves towards the pressure axis, the evolution appears to be slightly more complex with a tilt away from the scattering plane explored here. 

\begin{figure}
\includegraphics[width=0.95\linewidth,clip=]{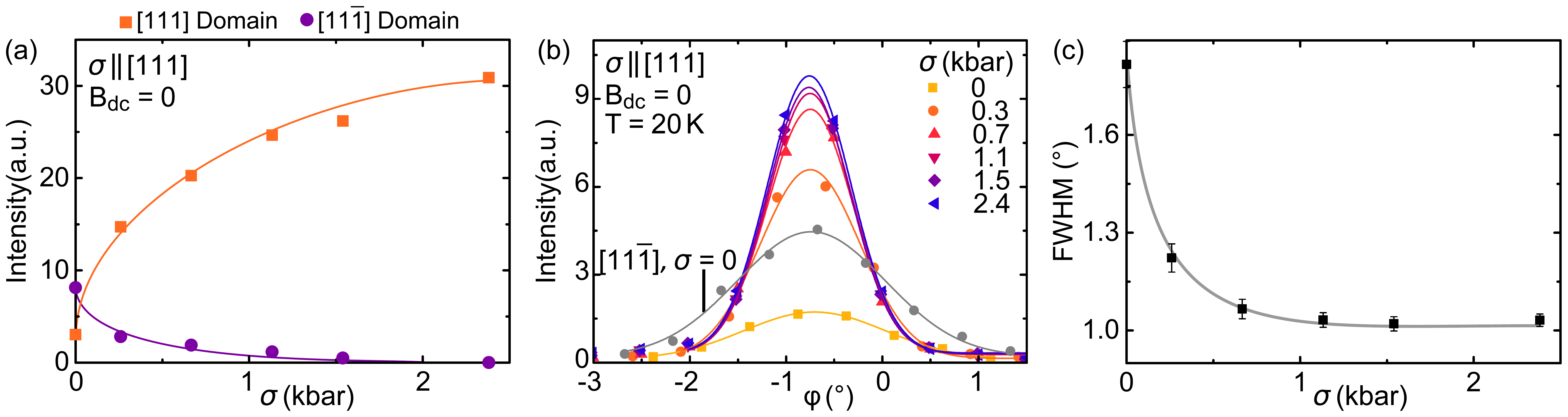}
\caption{(Colour online) Characteristics of the domain populations along \hkl [111] and \hkl [11-1], where uniaxial pressure was applied along \hkl [111] at zero field. (a) Peak intensities of the \hkl[111] and \hkl [11-1] domains as a function of pressure. (b) Intensity as a function of rocking angle $\phi$. Data of the \hkl [111] domain are shown for various pressures. For comparison data for the \hkl [11-1] domain are shown for $\vec{\sigma}=0$. (c) Full width half maximum (FWHM) of the rocking scans at zero field as a function of uniaxial pressure. At high pressures the FWHM approaches the resolution limit $\sim1^{\circ}$}
\label{Fig_S3}
\end{figure}

Fig.\,\ref{Fig_S3} summarises the evolution of the helical domain population along the \hkl [111] axis under uniaxial pressure parallel to \hkl [111].  We have confirmed that the pressure axis is parallel to within $1^{\circ}$ of the \hkl[111] axis. With increasing pressure the integrated intensity at zero field increases as shown in Fig.\,\ref{Fig_S3}\,(a). This reflects the increase of the domain population seen in Fig.\,\ref{Fig_S2}\,(a) through (c). We note that the \hkl[111] axis has a threefold symmetry. Thus the remaining three \hkl<111>  axes enclose the same angle with respect to the pressure axis. Being equivalent at zero field we assume that our observations for the domain along the \hkl[11-1] axis are also true for the other two domains that cannot be seen. As shown in Fig.\,\ref{Fig_S3}(a) the domain along the \hkl[11-1] axis depopulates quickly with increasing uniaxial pressure. Thus a single domain state may be generated by uniaxial pressure. Furthermore the full width at half maximum (FWHM) as obtained from rocking scans, cf. Fig.\,\ref{Fig_S3}\,(b), decreases rapidly and approaches the instrumental limit of $1^\circ$ as shown in Fig.\,\ref{Fig_S3}\,(c). Here the FWHM represents a measure of the magnetic mosaicity, where the decrease in mosaicity provides additional evidence that uniaxial pressure effectively increases the strength of the uniaxial magnetic anisotropy along the pressure axis.


\begin{figure}
\includegraphics[width=0.9\linewidth,clip=]{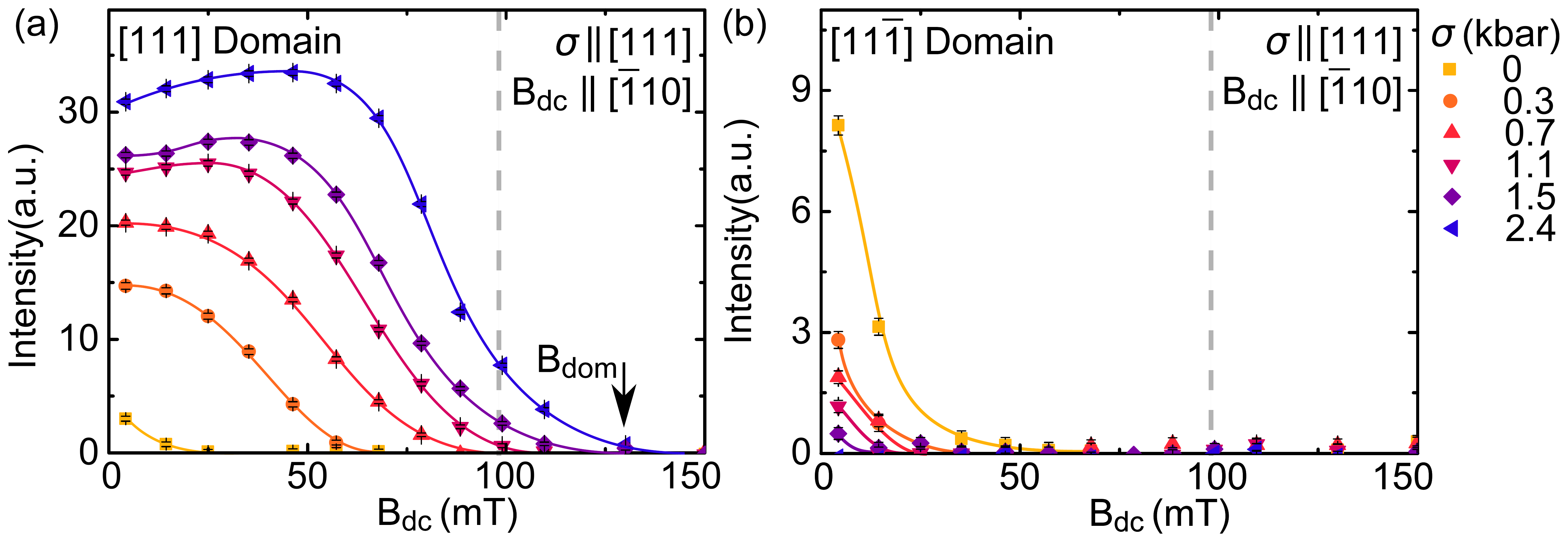}
\caption{(Colour online) Characteristics of the domain populations along \hkl [111] and \hkl [11-1], where uniaxial pressure was applied along \hkl [111]. (a) Peak intensity of the \hkl [111] domain (cf Fig.\,\ref{Fig_S2}) as a function of magnetic field \textit{perpendicular} to the pressure axis and parallel to \hkl [110] for various pressures. (b) Peak intensity of the \hkl [11 -1] domain (cf Fig.\,\ref{Fig_S2}) as a function of magnetic field for various pressures along \hkl [111]. The dash lines show the value of $B_{\rm{c1}}$ at ambient pressure.}
\label{Fig_S4}
\end{figure}

The three-fold symmetry along the \hkl<111> axes mentioned above is broken by a magnetic field applied perpendicular to the uniaxial pressure axis. At ambient pressure a field along a \hkl<110> axis leads to a depopulation of the helical domains along \hkl<111> axes perpendicular to the field, while the remaining two domains become populated. This depopulation takes place at a characteristic field $B_{\rm{dom}}\ll B_{\rm{c1}}$, cf. Fig.\,\ref{Fig_S4}\,(a). As the pressure increases the value of $B_{\rm dom}$ increases. This is not the case for the domain along the \hkl[11-1] direction, which instead becomes depopulated at smaller fields with increasing pressure, Fig.\,\ref{Fig_S4}\,(b). At high pressures the pressure induced uniaxial anisotropy becomes large and $B_{\rm dom}$ merges with the transition to the conical phase at $B_{\rm c1}$. Note that the data shown in Fig.\ref{Fig_S4} was measured at a temperature well below $T_{\rm{c1}}$ for increasing pressure to minimise any parasitic effects due to the decreasing transition temperature $T_{\rm{c1}}$.

\begin{figure}[h!]
\includegraphics[width=0.8\linewidth,clip=]{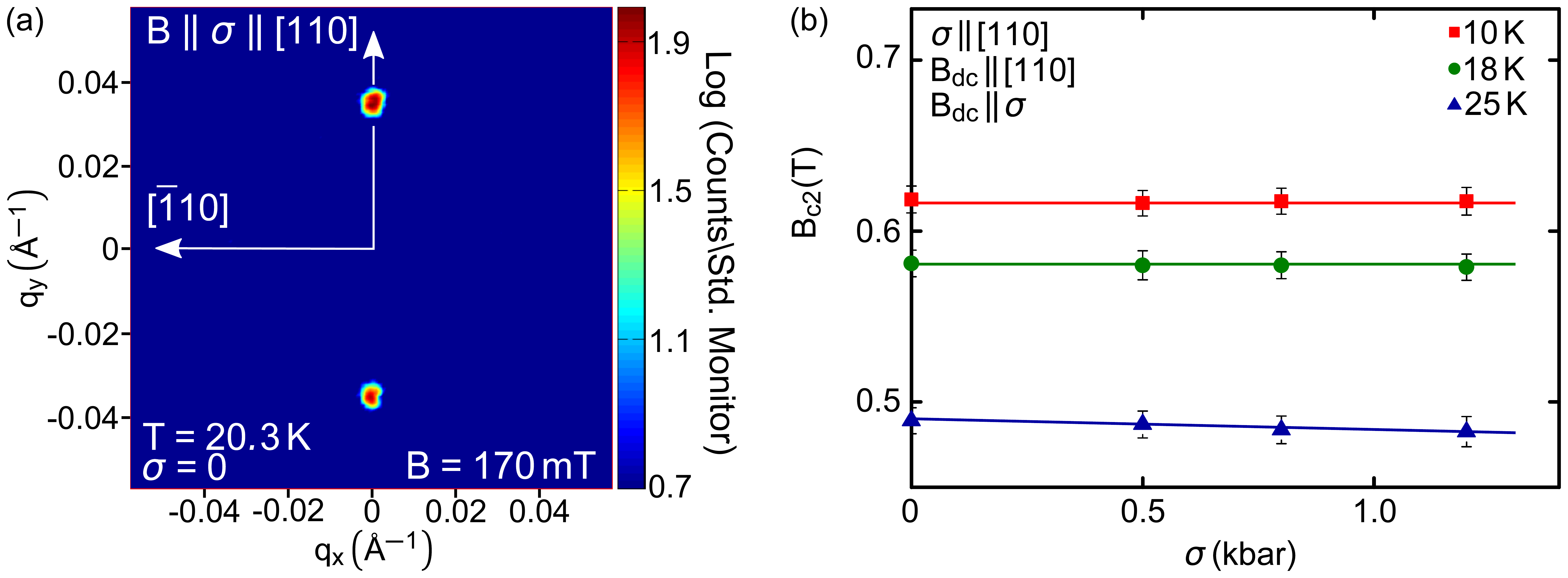}
\caption{(Colour online) Typical intensity pattern of the conical phase and pressure dependence of the upper critical field $B_{\rm c2}$ inferred from these data. (a) Intensity pattern in the conical phase for magnetic field parallel to the pressure axis (here the vertical direction), where field and pressure were applied along \hkl [110].
(b) Upper critical field $B_{\rm c2}$ as a function of pressure for selected temperatures. The pressure was applied along \hkl [110] parallel to the applied field.}
\label{Fig_S5}
\end{figure}

While the focus of our study was on the helical order and skyrmion lattice phase we have also collected data on the pressure dependence of the upper critical field $B_{\rm c2}$ for selected temperatures. Typical data are shown in Fig.\,\ref{Fig_S5}. As our main observation we do not find changes of $B_{\rm c2}$ under pressure. This is consistent with the pressure dependence of the $T_{\rm c}$, which agrees quantitatively with the hydrostatic pressure dependence of $T_{\rm c}$ for an isotropic material as discussed in the main text. Hydrostatic pressure experiments establish that $B_{\rm c2}$ is essentially unchanged up to closely below $p_{\rm c}=14.6\,{\rm kbar}$. In other words, the lack of a uniaxial pressure dependence of $B_{\rm c2}$ agrees with the unit cell volume dependence seen in $T_{\rm c}$.

\clearpage

\begin{figure}[h!]
\includegraphics[width=0.5\linewidth,clip=]{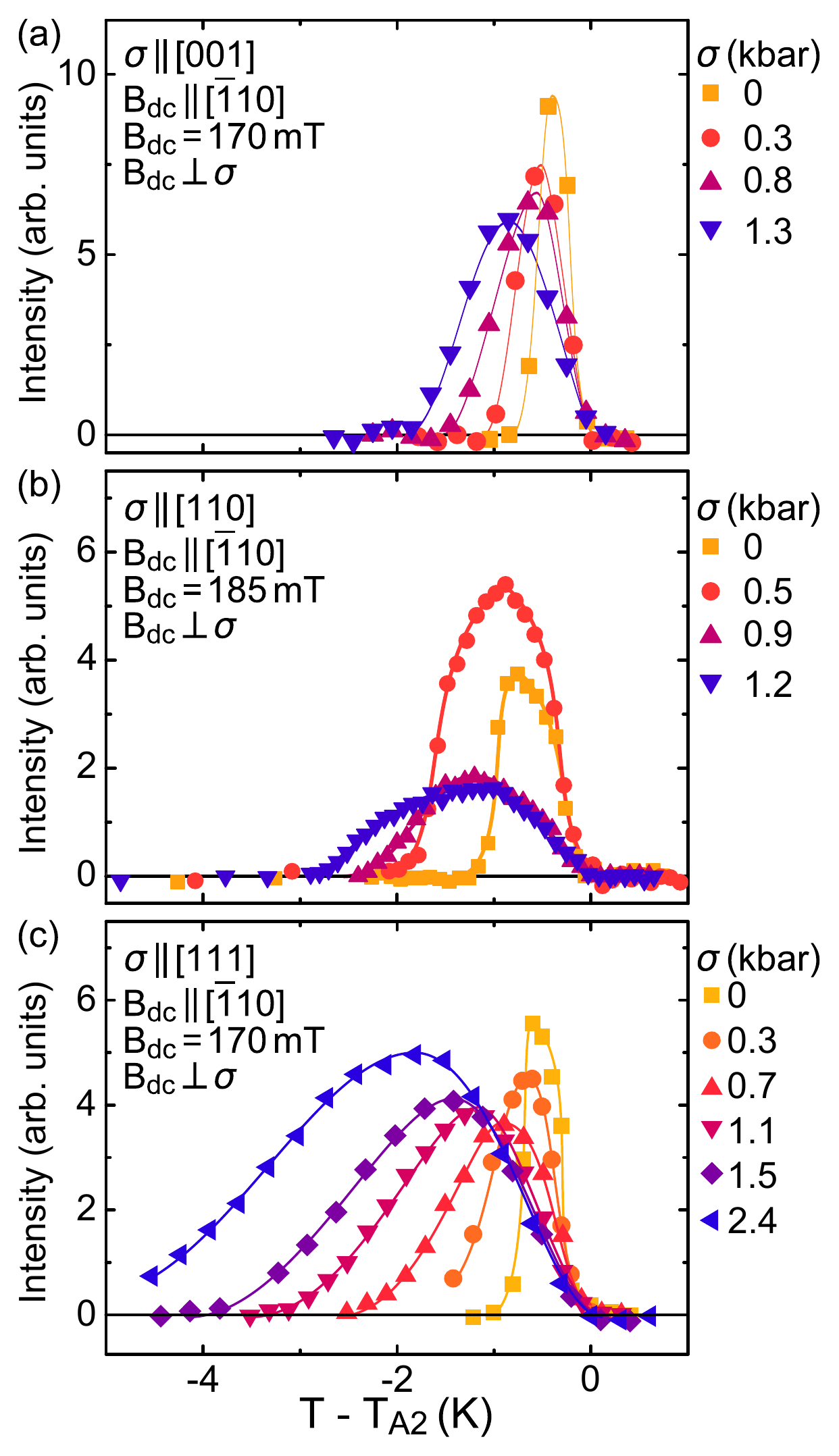}
\caption{(Colour online) Typical temperature dependencies of the peak intensities in the skyrmion lattice phase for various uniaxial pressures axes and magnetic field \textit{perpendicular} to the pressure axis. The magnetic field was always applied along \hkl [-110]. Temperatures are displayed with respect to $T_{\rm A2}$. (a) Peak intensities for pressures along \hkl [001]. (b) Peak intensities for pressures along \hkl [110]. (c) Peak intensities for pressures along \hkl [111]. }
\label{Fig_S6}
\end{figure}

\subsection{Skyrmion Lattice Phase}
 
Fig.\,\ref{Fig_S6} shows typical temperature dependencies of the SANS peak intensities in the skyrmion lattice phase for uniaxial pressures along \hkl [001], \hkl [110] and \hkl [111] and magnetic field \textit{perpendicular} to the pressure axis. For all pressure axes the magnetic field was always applied along \hkl [-110]. Temperatures are displayed with respect to the transition temperature to the paramagnetic state, $T_{\rm A2}$. Note that these measurements were carried out on different samples with the same dimensions as summarised in table\,\ref{table}. The field direction was chosen along \hkl [-110], which represents the only common crystallographic direction perpendicular to all three pressure directions. Values of the transition temperatures $T_{\rm A1}$ and $T_{\rm A2}$ reported in the main text were obtained from the data shown in Fig.\,\ref{Fig_S6}.


\begin{figure}
\includegraphics[width=0.7\linewidth,clip=]{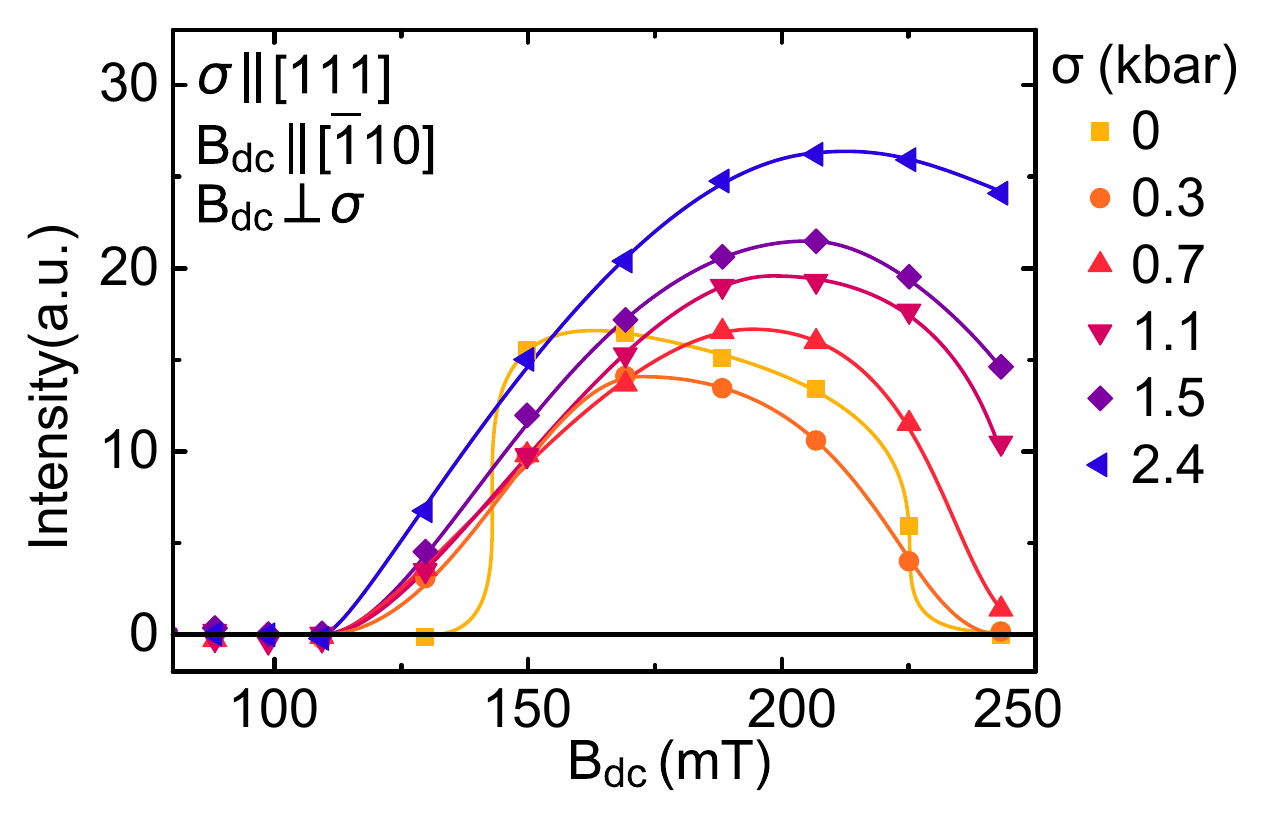}
\caption{(Colour online) Peak intensity in the skyrmion lattice phase for various uniaxial pressures along the  \hkl [111] axis as a function of magnetic field. Magnetic field was applied along \hkl[-110] perpendicular to the pressure axis. The magnet used for these measurements was limited to a maximum field of 250\,mT.}
\label{Fig_S7}
\end{figure}

Shown in Fig.\,\ref{Fig_S7} is the peak intensity in the skyrmion lattice phase for pressure along the  \hkl[111] axis as a function of magnetic field. These date are complementary to the data shown in Fig.\,\ref{Fig_S6}\,(c). Data were recorded at the temperature were maximum intensity was observed in the temperature scans shown in Fig.\,\ref{Fig_S6}\,(c) representing the location at which the skyrmion lattice phase is most stable. With increasing pressure the intensity increases. While the lower magnetic field, $B_{\rm A1}$, above which the skyrmion lattice phase appears, decreases only slightly for increasing pressure, our data suggest that the upper critical field of the skyrmion lattice phase, $B_{\rm A2}$, is strongly enhanced with increasing pressure. The magnet used for these measurements was limited to a maximum field of 250\,mT.


It is instructive to note, that the \hkl [110] axis represents an easy magnetic axis within the skyrmion lattice plane for all field directions except \hkl <100> \cite{muhlbauer2009skyrmion, adams2015, muenzer2010}. As stated in the main text, for uniaxial pressure along \hkl [110] and \hkl [001] the skyrmion lattice is aligned along \hkl [110] for all pressures. An additional second domain population, which is rotated by $30^{\circ}$ as compared to the first domain, emerges for high pressures along \hkl [110] as described in the main text. The orientation of the skyrmion lattice phase for pressures along \hkl [111] will be addressed further below.

\begin{figure}[h!]
\includegraphics[width=0.7\linewidth,clip=]{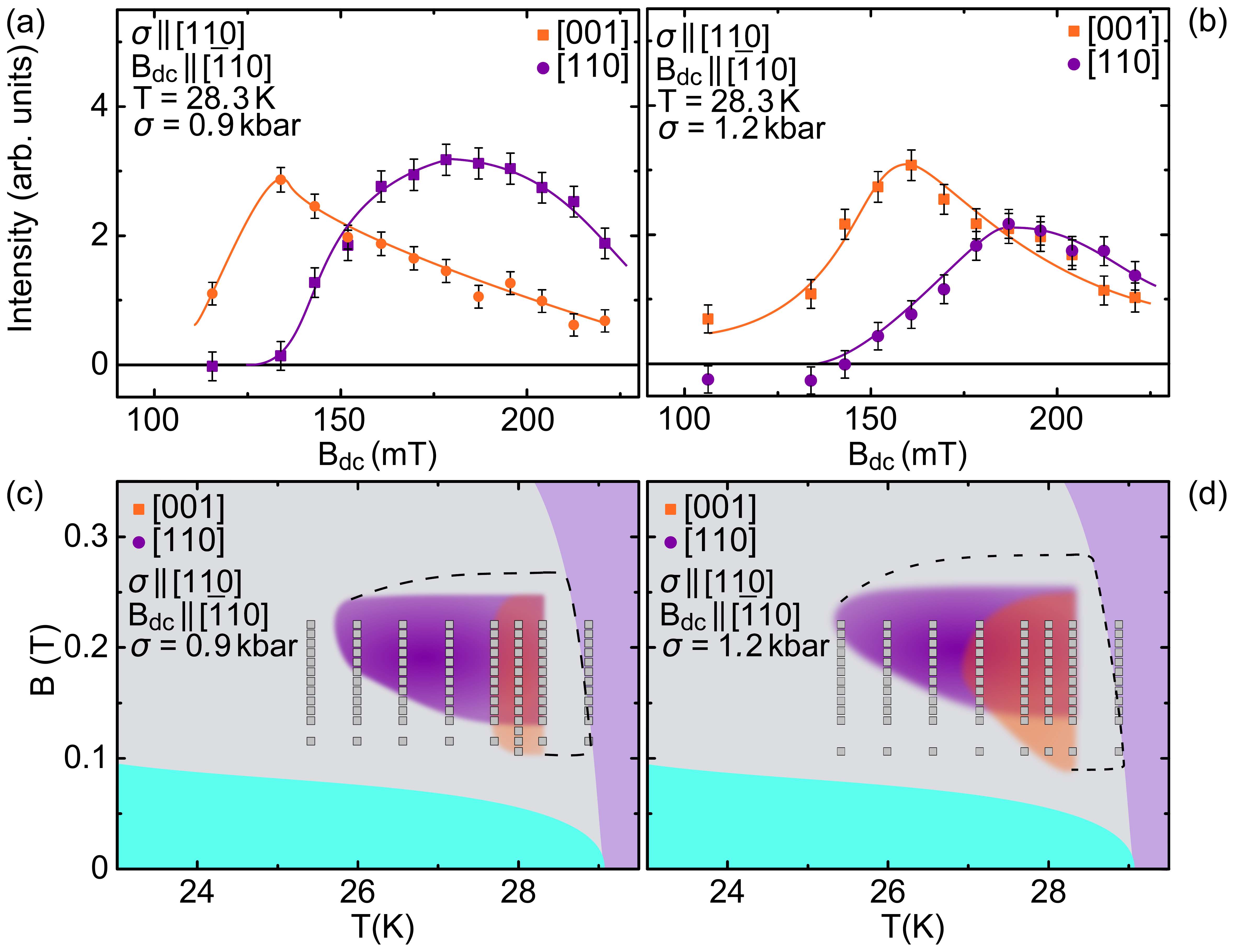}
\caption{(Colour online) Typical peak intensities as a function of magnetic field for pressures along \hkl [110] and magnetic phase diagrams inferred from these data. Panels (a) and (b): Peak intensities as a function of magnetic field along \hkl [-110] of the SLP domains along \hkl [110] and \hkl [001] under uniaxial pressure along \hkl [110]. Panels (c) and (d): Magnetic phase diagrams depicting the field and pressure regime where the different domain populations are observed. The grey data points mark the field and temperatures values at which data were recorded.}
\label{Fig_S8}
\end{figure}

At ambient pressure and magnetic field parallel to a \hkl <110> direction the SLP displays a single domain population oriented along a \hkl <110> direction in the plane perpendicular to the field. For pressures applied along \hkl [110] perpendicular to the applied field which is along \hkl [-110], i.e., in the plane of the SLP, a second domain population oriented along \hkl [001] is observed in field sweeps after zero field cooling. Typical intensity pattern are shown in Fig.\,1 in the main text. Typical data illustrating the temperature and magnetic field dependence of the two domain populations of the SLP for pressure along \hkl [110] are shown in Fig.\,\ref{Fig_S8}. The additional \hkl [001] oriented domain emerges at slightly smaller fields and coexist with the \hkl [110] oriented domain. The range in temperature and magnetic field in which the \hkl[001] oriented domain was observed increases with increasing pressure.

\begin{figure}[h!]
\includegraphics[width=0.8\linewidth,clip=]{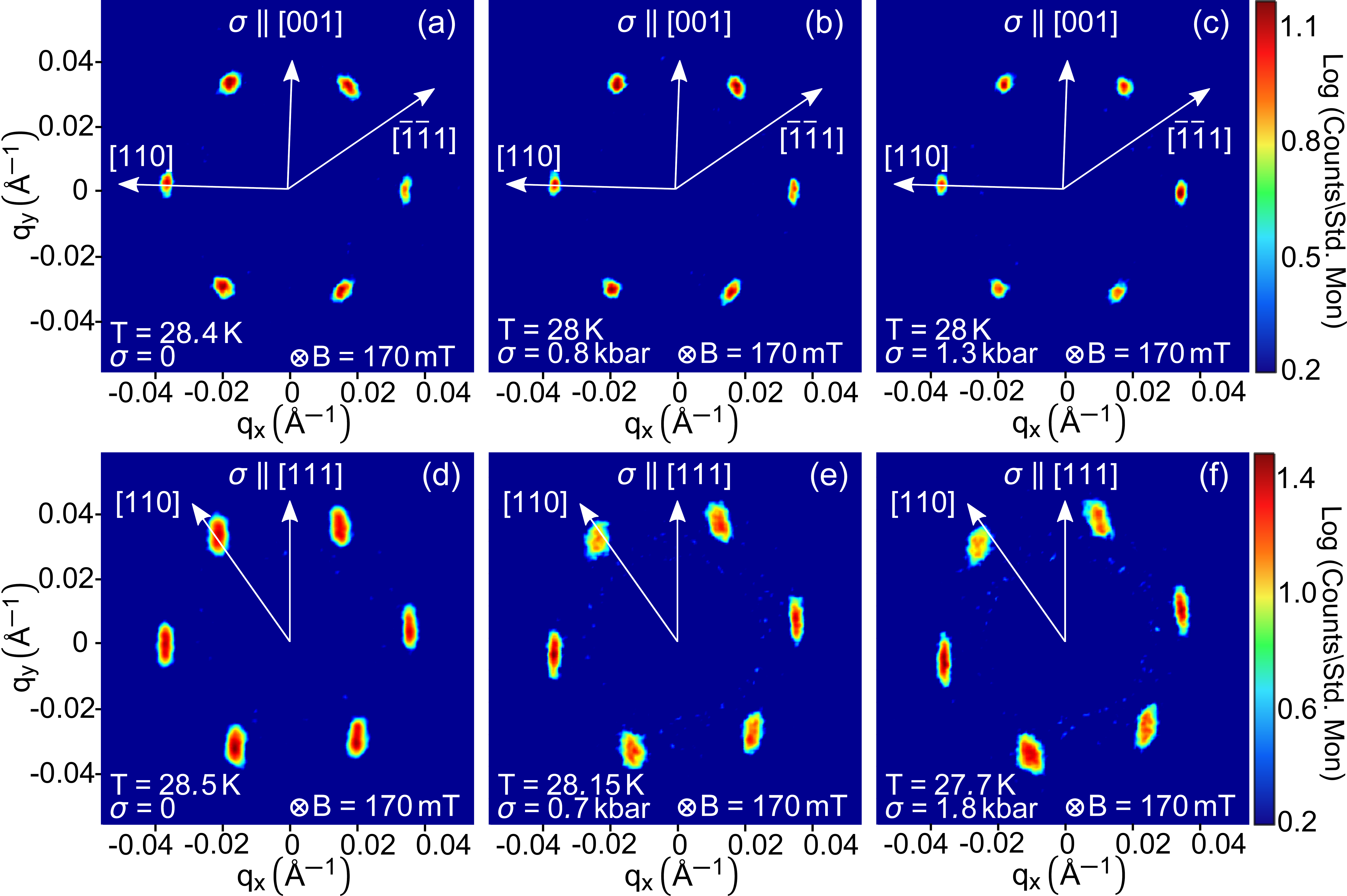}
\caption{(Colour online) Typical SANS intensity patterns in the skyrmion lattice phase for uniaxial pressure along \hkl [111] and \hkl [001]. Pressures were applied perpendicular to the applied field, which was along \hkl [110]. Panels (a) through (c): Intensity pattern for pressure along \hkl [001]. The intensity pattern is unchanged aligned along \hkl [110] for all pressures. There is no evidence suggesting the formation of a second domain. Panels (d) through (f): Intensity pattern for pressure along \hkl [111] and different temperatures. At low pressure and temperature near $T_{\rm A2}$ the SLP is aligned along \hkl [110]; with increasing pressure and decreasing temperature the pattern rotates towards the pressure axis. }
\label{Fig_S9}
\end{figure}

Shown in Fig.\,\ref{Fig_S9} are typical SANS intensity patterns in the skyrmion lattice phase for uniaxial pressure along \hkl [001] and \hkl [111] perpendicular to an applied field along \hkl [-110].  Shown in Fig.\,\ref{Fig_S9}\,(a) through (c) are typical intensity patterns for pressure along \hkl [001]. Here the intensity pattern remains unchanged aligned along \hkl [110] for all pressures. There is no evidence for the formation of a second domain as observed for pressure along \hkl [110]. For small pressures along \hkl [111] and temperatures near $T_{\rm A2}$ the SLP is aligned along \hkl [110] as shown in Fig.\,\ref{Fig_S9}\,(d). For large pressures along \hkl [111] and decreasing temperatures the pattern rotates by an angle $\omega$ towards the pressure axis as shown in Fig.\,\ref{Fig_S9}\,(e) and (f).

\begin{figure}
\includegraphics[width=0.95\linewidth,clip=]{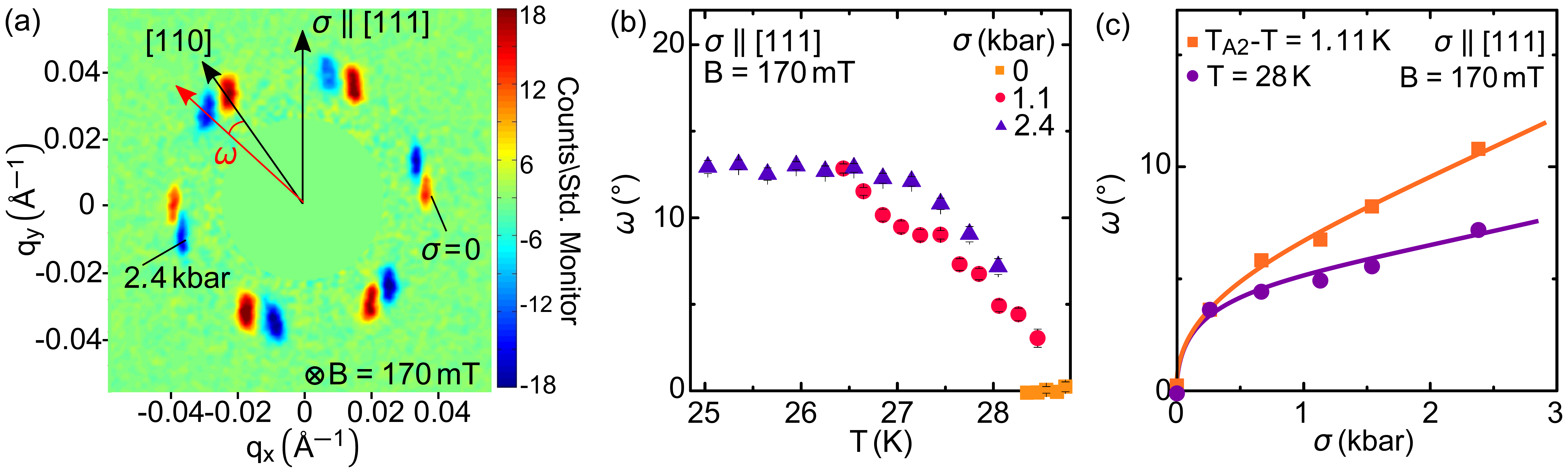}
\caption{(Colour online) Illustration of changes of the SANS intensity pattern in the SLP for uniaxial pressure along \hkl [111] and magnetic field along \hkl [-110] and perpendicular to the pressure axis. The pressure axis is along the vertical direction.  (a) Resulting scattering pattern after substracting the data at $\sigma\!=\!2.4$\,kbar and $T\!=\!26$\,K (blue minima) from the data obtained at $\sigma\!=\!0$ and $T\!=\!28.5$\,K (red maxima), both measured at $B\!=\!170$\,mT. With increasing pressure the pattern rotates towards the pressure axis. (b) Rotation angle $\omega$ between the intensity pattern and the pressure axis as defined in panel\,(a). At high temperatures the pattern is aligned along \hkl [110] with decreasing temperature it rotates towards the pressure axis. (c) Rotation angle $\omega$ as a function of pressure for a temperature of 28\,K and a field of 170\,mT.}
\label{Fig_S10}
\end{figure}

This rotation is made apparent in Fig.\,\ref{Fig_S10}\,(a), where the scattering pattern obtained at $\sigma\!=\!2.4\,\rm{kbar}$ and $T\!=\!26\,\rm{K}$ is subtracted from the scattering pattern obtained at ambient pressure at $T\!=\!28.5\,\rm{K}$. The resulting scattering pattern contains blue minima representing the bragg peaks position under pressure, while the bragg spots at ambient pressure are given by red maxima. The magnitude of $\omega$ is a function of both temperature and uniaxial pressure as shown on Fig.\,\ref{Fig_S10}\,(b) and (c). At ambient pressure $\omega$ shows no temperature dependence. With increasing temperature the rotation angle decreases suggesting that thermal fluctuations suppress the rotation. The evolution of $\omega$ as a function of uniaxial pressure is obtained by extracting its value at a fixed constant temperature of 28\,K and different pressures as shown in Fig.\,\ref{Fig_S10}\,(c). Since $T_{\rm{A2}}$ changes as a function of pressure we also show for completeness the values of $\omega$ for a constant temperature difference of 1.1\,K below $T_{\rm{A2}}$.

\begin{figure} [h!]
\includegraphics[width=0.95\linewidth,clip=]{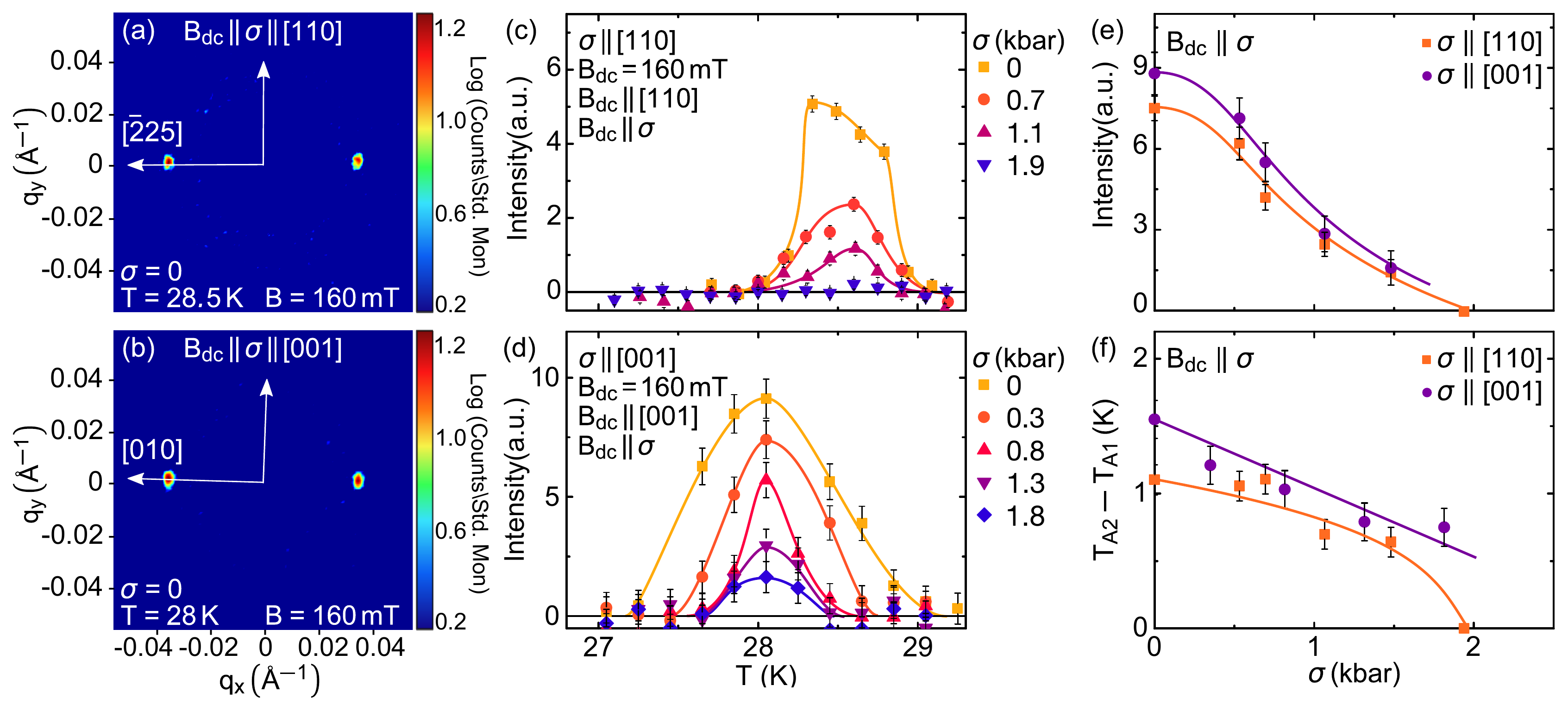}
\caption{(Colour online) Typical scattering patterns and temperature dependencies of the peak intensities in the skyrmion lattice phase for various uniaxial pressures axes and magnetic field \textit{parallel} to the pressure axis. (a) Typical scattering pattern of the SLP for magnetic field and uniaxial pressure along the \hkl[110] crystal axis. (b) Typical scattering pattern of the SLP for magnetic field and uniaxial pressure along the \hkl[001] crystal axis, note the change in orientation of the SLP from \hkl<110> to \hkl<100> (c) Peak intensities for pressures along \hkl [110]. (d) Peak intensities for pressures along \hkl [001]. (e) Maximum peak intensity within the skyrmion lattice as a function of uniaxial pressure for magnetic field \textit{parallel} to the pressure axis. (f) Temperature width of the SLP as a function of uniaxial pressure for \textit{parallel} to the pressure axis.}  
\label{Fig_S11}
\end{figure}

Fig.\,\ref{Fig_S11} shows typical SANS scattering patterns for uniaxial pressure along \hkl [001] and \hkl [110] and magnetic field \textit{parallel} to the pressure axis. We note, that for magnetic field perpendicular to the incoming neutron beam only two of the six Bragg peaks of the skyrmion lattice phase may be observed simultaneously, c.f. Fig.\,\ref{Fig_S1}. In the case of uniaxial pressure and magnetic field \textit{parallel} to the \hkl [110] axis the helical order may be observed when a \hkl [-110] is parallel to the neutron beam. In order to measure the intensity from the skyrmion lattice the sample must therefore be rotated by $30^\circ$ around the vetical \hkl [110] axis. The intensity observed is then along a \hkl[-225] direction, which is $60^\circ$ away from \hkl [-110], cf Fig.\,\ref{Fig_S11}\,{(a)}.

As mentioned before, due to the symmetries of the B20 structure the \hkl<110> direction does no represent an easy axis for the skyrmion lattice phase under magnetic fields applied along a \hkl <100> axis \cite{muenzer2010}. Instead the skyrmion lattice rotates by $15^\circ$ within the plane so that one of the \textbf{Q} vectors points along a \hkl<100> perpendicular to the field, cf. Fig.\,\ref{Fig_S11}\,(b). Fig.\,\ref{Fig_S11}\,(c) and (d) show the peak intensity as a function of temperature for different uniaxial pressures. With increasing pressure the peak intensity decreases and vanishes at the largest pressures studied, while the temperature range in which the skyrmion lattice phase is stable decreases with increasing uniaxial pressure. Taken together this provides further evidence that the application of uniaxial pressure parallel to the applied magnetic field causes a rapid suppression of the skyrmion lattice phase.


